\documentclass[iop]{emulateapj}
\usepackage{graphicx,natbib,apjfonts}
\usepackage{arydshln}
\usepackage{amsmath}

\begin{document}

\submitted{Accepted October 9, 2014}

\title{The Clustering and Halo Masses of Star Forming Galaxies at $\lowercase{z}<1$}
\author{Tim Dolley${}^{1,2}$}
\author{Michael J.~I.~Brown${}^{1,2}$}
\author{Benjamin J.~Weiner${}^3$}
\author{Mark Brodwin${}^4$}
\author{C.~S.~Kochanek${}^{5,6}$}
\author{Kevin A.~Pimbblet${}^{1,2,7}$}
\author{David P.~Palamara${}^{1,2}$}
\author{Buell T.~Jannuzi${}^3$}
\author{Arjun Dey${}^{8,9}$}
\author{David W.~Atlee${}^8$}
\author{Richard Beare${}^{1,2}$}

\affiliation{${}^1$School of Physics and Astronomy, Monash University, Clayton, Victoria 3800, Australia}
\affiliation{${}^2$Monash Centre for Astrophysics (MoCA), Monash University, Clayton, Victoria 3800, Australia}
\affiliation{${}^3$Steward Observatory, University of Arizona, Tucson, AZ 85721, USA}
\affiliation{${}^4$Department of Physics and Astronomy, University of Missouri, Kansas City, MO 64110, USA}
\affiliation{${}^5$Department of Astronomy, The Ohio State University, Columbus, OH 43210, USA}
\affiliation{${}^6$Center for Cosmology and Astroparticle Physics, The Ohio State University, Columbus, OH 43210, USA}
\affiliation{${}^7$Department of Physics and Mathematics, University of Hull, Kingston-upon-Hull, HU6 7RX, UK}
\affiliation{${}^8$National Optical Astronomy Observatory, Tucson, AZ 85719, USA}
\affiliation{${}^9$Radcliffe Institute for Advanced Study, Cambridge, MA 02138, USA}

\email{Tim.Dolley@monash.edu}

\begin{abstract}

We present clustering measurements and halo masses of star forming galaxies at $0.2<z<1.0$.  After excluding AGN, we construct a sample of 22553 $24~\mu$m sources selected from $8.42\rm{~deg}^2$ of the \emph{Spitzer} MIPS AGN and Galaxy Evolution Survey of Bo\"otes.  Mid-infrared imaging allows us to observe galaxies with the highest star formation rates (SFRs), less biased by dust obscuration afflicting the optical bands.  We find that the galaxies with the highest SFRs have optical colors which are redder than typical blue cloud galaxies, with many residing within the green valley.  At $z>0.4$ our sample is dominated by luminous infrared galaxies (LIRGs, $L_{TIR}>10^{11}~L_\odot$) and is comprised entirely of LIRGs and ultra-luminous infrared galaxies (ULIRGs, $L_{TIR}>10^{12}~L_\odot$) at $z>0.6$.  We observe weak clustering of $r_0\approx3-6~h^{-1}$Mpc for almost all of our star forming samples.  We find that the clustering and halo mass depend on $L_{TIR}$ at all redshifts, where galaxies with higher $L_{TIR}$ (hence higher SFRs) have stronger clustering.  Galaxies with the highest SFRs at each redshift typically reside within dark matter halos of $M_{halo}\approx10^{12.9}~h^{-1}M_\odot$.  This is consistent with a transitional halo mass, above which star formation is largely truncated, although we cannot exclude that ULIRGs reside within higher mass halos.  By modeling the clustering evolution of halos, we connect our star forming galaxy samples to their local descendants.  Most star forming galaxies at $z<1.0$ are the progenitors of $L\lesssim2.5L_*$ blue galaxies in the local universe, but star forming galaxies with the highest SFRs ($L_{TIR}\gtrsim10^{11.7}~L_\odot$) at $0.6<z<1.0$ are the progenitors of early-type galaxies in denser group environments.
\end{abstract}

\maketitle

\section{Introduction}

The observed color distribution of galaxies is bimodal \citep[e.g.][]{strat01,baldr04,hogg04}: most red galaxies contain little gas and dust and are comprised of old red stars which formed at high redshift, while blue galaxies are undergoing active star formation.  Star formation occurs when gas cools and collapses under the influence of gravity, but what stops new stars from forming?  Models of galaxy formation are able to reproduce the observed bimodal color distribution, with various mechanisms such as virial shock heating, active galactic nuclei (AGN), supernovae feedback, and starbursts caused by galaxy interactions \citep[e.g.][]{bir03,menci05,ker05,croto06,menci06,dek06,bir07,dre09, cen11}, but the contribution from each of these mechanisms is uncertain.

Furthermore, the stellar mass in galaxies grows by a combination of star formation and stars acquired through mergers, but it is still unclear which process dominates, and how it depends on galaxy environment. The most massive red galaxies are thought to grow to such sizes via mergers, since we do not see star forming galaxies of such mass in the local universe.  However, there is observational evidence for the existence of massive star forming galaxies at $z>1$ \citep[e.g.][]{glaze04,sha04,farra06,brodw08,brodw13,magli08,palam13,lee13}, so it is plausible that massive red galaxies are formed by the truncation of star formation in massive blue galaxies.  Understanding the history and evolution of star forming galaxies is necessary to explain the observed abundances and morphologies of galaxies we see today.

By measuring the clustering of star forming galaxies at various epochs, we can determine their mean dark matter halo masses \citep[e.g.][]{selja00,smith03,zhe05,zhe09,coil08,zehav11}.  This lets us know the typical environment that star forming galaxies reside within, which can be used to distinguish between various modes of star formation and to constrain the mechanisms responsible for the truncation of star formation.  For example, if star formation is truncated as galaxies fall into massive halos, then low mass red galaxies will have enhanced clustering compared to other galaxies of comparable mass, and for merger driven star formation we would expect an excess in clustering at small scales.  Since dark matter only interacts gravitationally, the space density and clustering of dark matter halos are predictable functions of redshift \citep[e.g.][]{selja00,sprin05}.  This allows us to use galaxy clustering to connect distant galaxy populations to today{\textquoteright}s galaxies in an evolutionary sequence.

Star forming galaxy samples are typically selected based on their optical color, since morphologies are only well determined for low redshift galaxies.  Unfortunately, dust obscuration can heavily bias the optical colors of galaxies, excluding many star forming galaxies from optically selected samples.  Emission in the mid-infrared (MIR) is primarily from dust heated by young hot stars \citep[e.g.][]{elbaz02,bell03,leflo05,desai08,lacey08,magli08,treye10}.  This provides us with samples of galaxies with high star formation rates (SFRs) that are less biased by the varying dust obscuration afflicting optical selection of star forming galaxies.  So by measuring the clustering of $24~\mu$m sources, we can examine how the environment of star forming galaxies has changed over cosmic time and connect star forming galaxies from earlier epochs to their descendants in the local universe to see how these star forming galaxies have evolved.

Previous MIR clustering results suffer from several limitations created by using small fields and small samples.  Small fields may not contain representative populations of galaxies due to cosmic variance.  For example, the variance in the galaxy number density for each of the $\sim160~\rm{arcmin}^2$ GOODS fields between $0.8<z<1.0$ is approximately $30\%$ \citep{somer04,drive10}.  Small samples also result in low pair counts and uncertain clustering measurements, often with underestimated uncertainties.  To alleviate these problems, galaxies from large redshift ranges are often grouped together \citep[e.g.][]{magli08,gilli07,stari12}, producing clustering results from combined populations of star forming galaxies over a broad range of cosmic history, which may not be indicative of any one of the individual populations, and also hinders the ability to measure clustering evolution.  Higher redshift samples generally lack spectroscopic redshifts, so rely on approximations of the redshift distribution to infer the spatial clustering.  Even if photometric redshifts are available, these often have large uncertainties which increase with redshift, and if these uncertainties are not treated correctly, they can greatly affect the measured spatial clustering.

Previous studies of MIR clustering have found relatively low clustering strengths at $z<1$, with correlation lengths ranging from $r_0=3.4-6.0~h^{-1}$Mpc \citep{fishe94a,magli07,magli08,gilli07,stari12}.  This is similar to that of blue galaxy samples, as measured by \citet{zehav11} at $z\simeq0.1$ and by \citet{coil08} at $z\simeq1$.  Higher redshift studies ($1<z<3$) find larger correlation lengths, up to $14.4~h^{-1}$Mpc \citep{farra06,magli07,magli08,brodw08,stari12}, which increase with redshift.  This strong clustering is much higher than the $r_0=5.02\pm0.07~h^{-1}$Mpc found for the most luminous local blue galaxy samples \citep{zehav11}.   While these MIR samples suffer from many of the previously mentioned limitations, they suggest a rapid change in the environment of star forming galaxies has occurred since $z=3$.  This indicates that star formation occurred in more massive galaxies lying in denser environments at high redshifts.

In this paper we present the most robust clustering measurements of $24~\mu$m sources to date for the redshift range $0.2<z<1.0$.  We determine the dark matter halo masses of star forming galaxies over this time, and connect them to their descendant galaxy populations.  We adopt a flat $\Lambda$CDM cosmology with $\Omega_m=0.27$, $\Omega_b=0.045$, $h=0.704$, $\sigma_8=0.81$ and $n_s=0.96$, consistent with the \citet{komat11} 7-year WMAP results.  All distances are given in comoving coordinates and Vega magnitudes are used throughout.  \emph{Spitzer} IRAC and MIPS apparent magnitudes are quoted using square brackets (e.g. $3.6~\mu$m is $[3.6]$).

\section{Data}

\subsection{The 24$~\mu$\lowercase{m} Sample}

The $24~\mu$m galaxy sample was selected from $8.42~{\rm{~deg}^2}$ of the \emph{Spitzer} MIPS AGN and Galaxy Evolution Survey \citep[MAGES;][Jannuzi et al.\ in prep.]{jannu10} of the Bo\"otes field, which overlaps the optical ($B_WRI$) NOAO Deep Wide Field Survey \citep[NDWFS;][]{jannu99}.  We also used the MIR photometry ($3.6-8.0~\mu$m) from the \emph{Spitzer} IRAC Deep Wide Field Survey \citep[SDWFS;][]{ashby09}.  We used spectra from the AGN and Galaxy Evolution Survey \citep[AGES;][]{kocha12} and some spectra from various smaller projects from Gemini and Keck to produce and analyse the uncertainties in our photometric redshifts and to quantify the effectiveness of our color selections.

\begin{figure*}
\begin{center} 
\includegraphics[width=1.8\columnwidth]{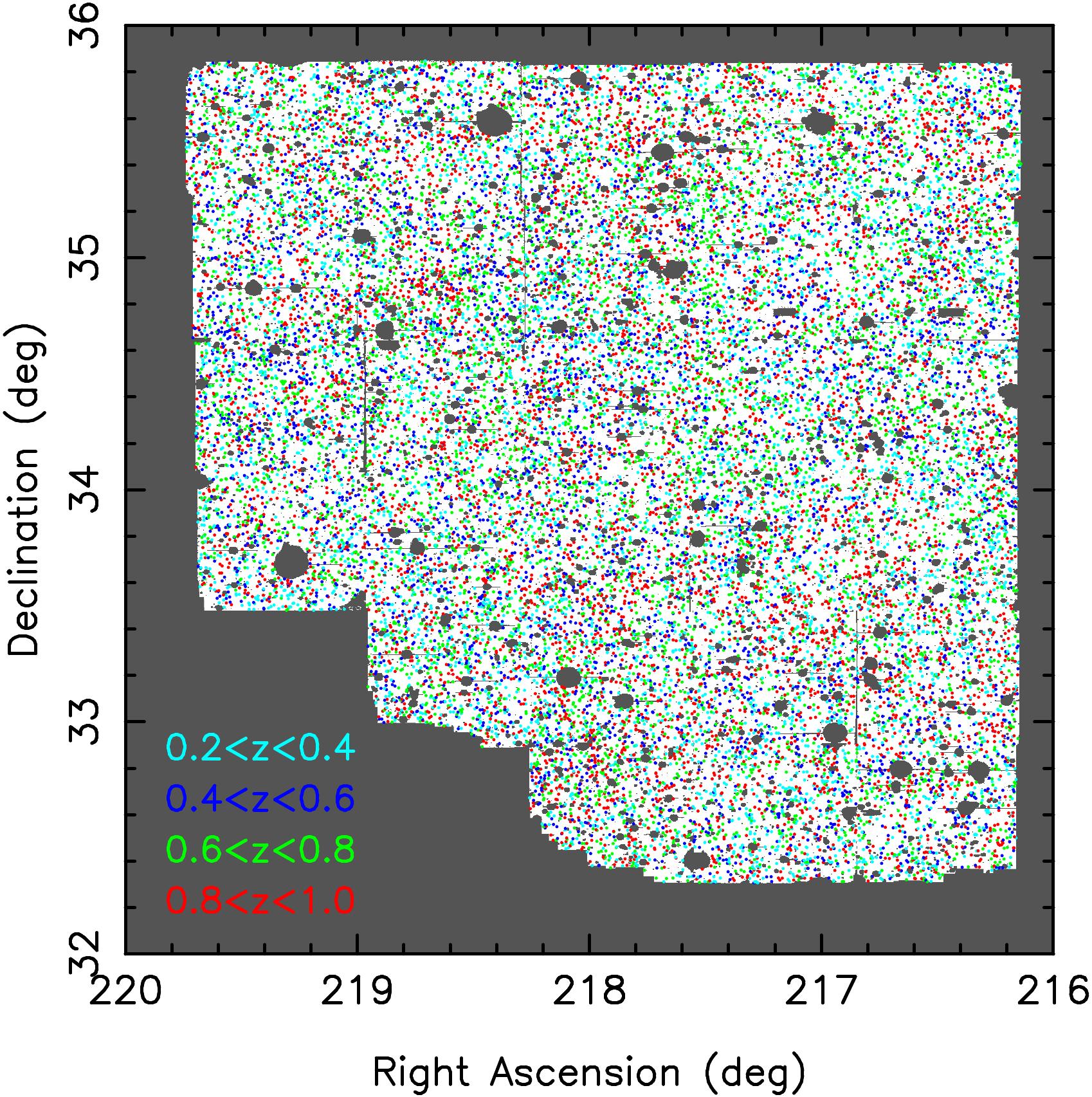}
\caption{Our star forming galaxy sample selected from $8.42\rm{~deg}^2$ of the Bo\"otes field.  Galaxies are colored by redshift.  The grey region shows the mask for the field, where galaxies were removed due to contamination by bright foreground objects or a lack of coverage in any band.}
\label{fig:gals}
\end{center}
\end{figure*}

MAGES sources were matched to $I$-band selected NDWFS sources, allowing for astrometric offsets. The $I$-band source catalog was created with SExtractor \citep{ber96} in single-image mode.  For further details about the photometry and catalog generation we refer the reader to \citet{brown08} and \citet{jannu99}.  A region mask was created for the field, where galaxies were removed from the sample.  The mask and sample distribution are shown in Figure \ref{fig:gals}.  Masked regions are due to contamination by bright foreground objects or the lack of coverage in any band.

The final sample was restricted to photometric redshifts of $0.2<z<1.0$, due to the larger uncertainties in photometric redshifts beyond this range.  We used a flux limit of $F_{24~\mu\rm{m}} > 0.223$~mJy, which is the $5\sigma$ detection limit of the MAGES catalog.  The sample completeness at this flux limit is greater than $95\%$.  After star and AGN removal (Section \ref{sec:starsAGN}) the final sample contains 22553 sources, of which 5089 have spectroscopic redshifts.  We split the sample into redshift limited subsamples to examine the evolution of the clustering of star forming galaxies.  We also use a second sample of 7799, $F_{24~\mu\rm{m}} > 0.4$~mJy sources for direct comparisons with previous $24~\mu$m clustering measurements and to examine any luminosity dependence.

\subsection{Star and AGN Removal}
\label{sec:starsAGN}

\begin{figure*}
\begin{center} 
\resizebox{2\columnwidth}{!}{\plottwo{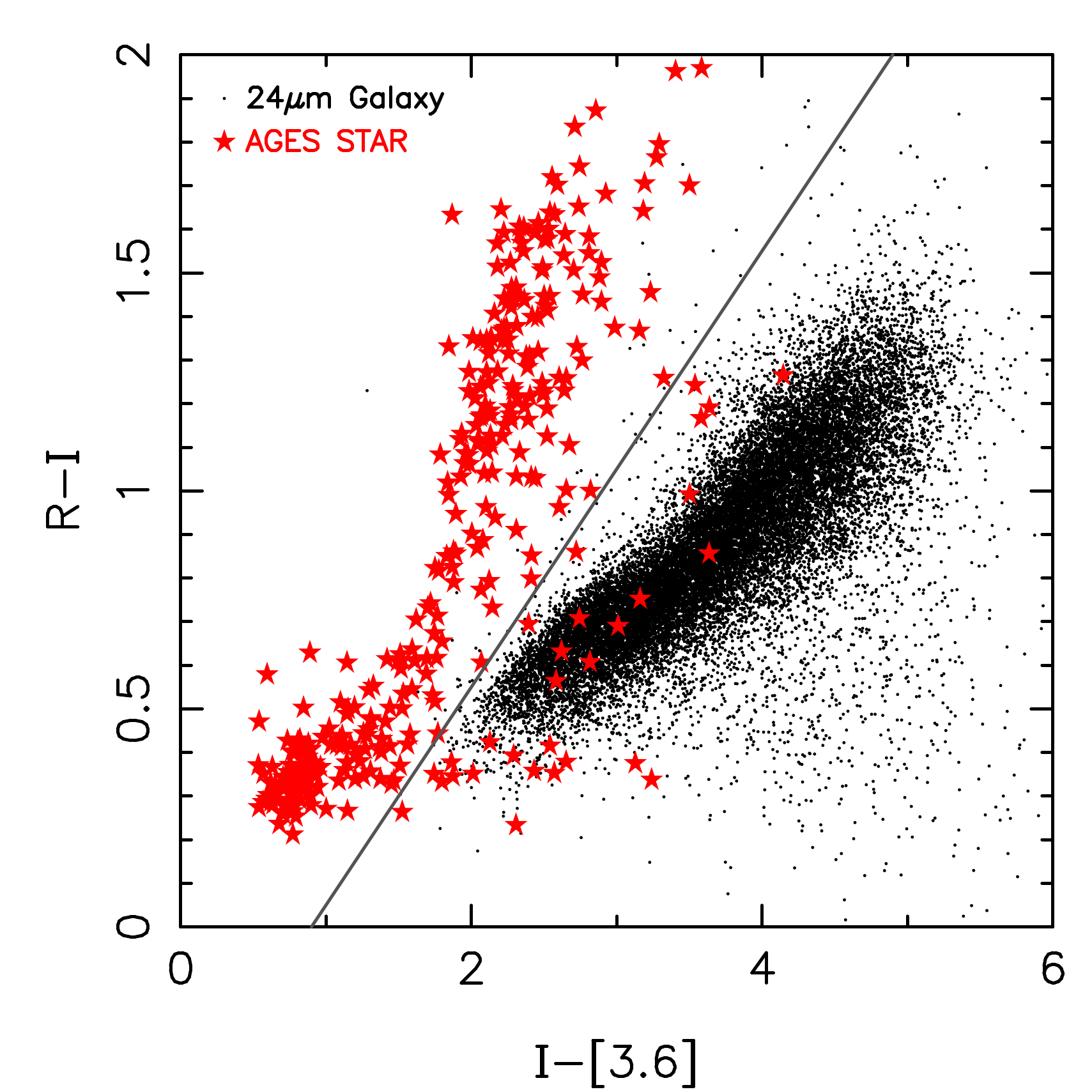}{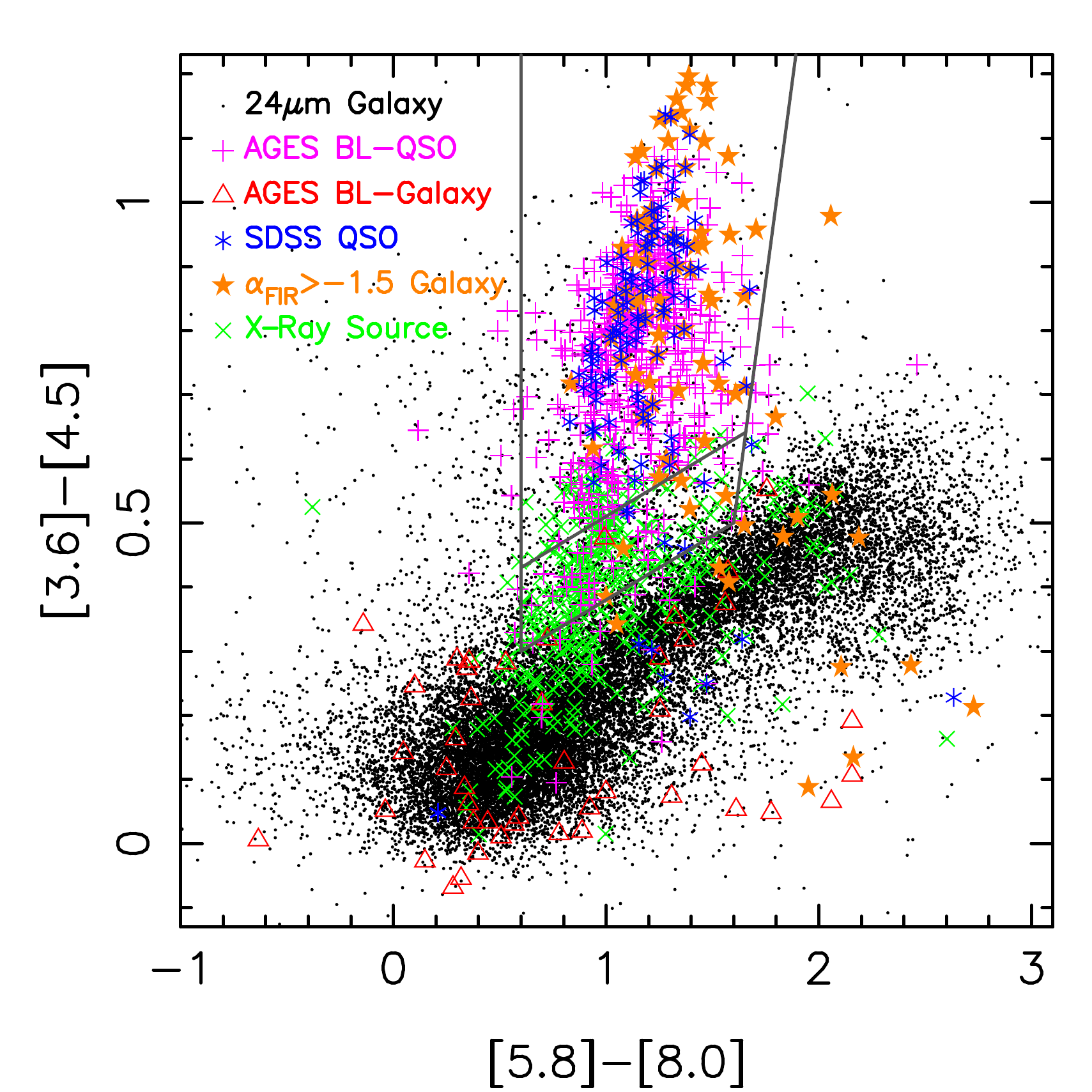}}
\caption{Color cuts used to remove stars (left) and AGN (right) from the $24~\mu$m sample. Left: Spectroscopically identified stars from AGES are shown by the star symbols.  Stars have a deficiency in $3.6~\mu$m flux compared to galaxies and are easily separated.  Right: Quasars identified by AGES and SDSS spectra, XBo\"otes X-ray sources, and from hot dust emission ($\alpha_{FIR}>-1.5$).  Our modified S05 AGN removal cut removes over $80\%$ of known quasars in the field, but removes fewer valid star forming galaxies from our sample.}
\label{fig:colcuts}
\end{center}
\end{figure*}

Foreground stars were removed with the color cut $R-I < 0.5(I - [3.6]) - 0.4$ (Figure \ref{fig:colcuts}), based on \citet{brown07}.  This cut removed 306 (89\%) of the AGES spectroscopically identified stars in our field and 67 (0.29\%) objects from our $24~\mu$m sample.  Even without this cut, stellar contamination is unlikely to be a concern.  Most stars are below our $24~\mu$m flux limit, since this is sampling the Rayleigh-Jeans tail of stellar blackbody emission.  Less than $1\%$ of AGES spectroscopically identified stars had $24~\mu$m detections above our flux limit.

AGN were removed from the sample using a modification of the \citet{stern05} criteria for AGN selection (hereafter S05).  The S05 mid-infrared color cut was found to remove too many $24~\mu$m sources believed to be star forming galaxies, particularly at $z\approx0.5$ where the galaxy locus moves into the S05 AGN region.  \citet{donle08} show that tracks of star forming SEDs cross the S05 region from $0.2\lesssim z\lesssim1.0$ but predominantly at $z\approx0.5$.  Figure \ref{fig:colcuts} shows our modification to the base line of the S05 cut, which was moved to $([3.6]-[4.5])=0.2([5.8]-[8.0])+0.31$.  This less aggressive AGN exclusion cut removed 382 (1.6\%) objects from our sample.   If we use the standard S05 cut to remove AGN, the measured clustering changes by less than $1\sigma$ for all samples.

We examined the effectiveness of our modified S05 AGN removal cut with known AGN in the field. It was found to remove 390 (84\%) of the spectroscopically identified quasars in AGES, and 109 (91\%) of the spectroscopically identified Sloan Digital Sky Survey (SDSS) quasars within our field.  \citet{mauch07} show that AGN typically have a FIR spectral index of $\alpha_{FIR}>-1.5$, as their central engine heats dust to higher temperatures than star formation.  For all MAGES objects which also had a $70~\mu$m detection, we find that our AGN cut removes 68 (77\%) galaxies with $\alpha_{FIR}=\log(F_{24}/F_{70})/\log(\nu_{24}/\nu_{70})>-1.5$.

While the S05 color cut removes quasars from the AGES sample, it does not remove all Seyferts (including broad line Seyfert I galaxies) where stellar emission dominates the spectral energy distribution.  We remove known X-ray sources from our sample by cross-matching with the XBo\"otes point source catalog \citep{ken05}.  This removed 376 (1.6\%) objects from our sample.  \citet{bra06} show that the majority of point sources in XBo\"otes are AGN.  As shown in Figure \ref{fig:colcuts}, these sources predominantly lie in a tight region at the base of the S05 AGN locus, indicating that their AGN are contributing significantly to their MIR emission.  Leaving these sources in the sample has almost no effect on the measured clustering.  Some Seyferts may still be left in our star forming galaxy sample, but as they are not removed by the S05 cut and not found in the XBo\"otes catalog, it is likely that their MIR emission is dominated by dust heated by star formation, so they are correctly assigned photometric redshifts, and should be kept in the sample if they have a $24~\mu$m detection. Also, broad line Seyferts are less than 1\% of all AGES $24\mu$m sources, so even if the AGN are contributing to the $24~\mu$m emission from these galaxies, they are a small fraction of the sample, and unlikely to have a significant effect on clustering measurements.

\subsection{Photometric Redshifts}
\label{sec:zphot}

\begin{figure}
\begin{center}
\includegraphics[width=\columnwidth]{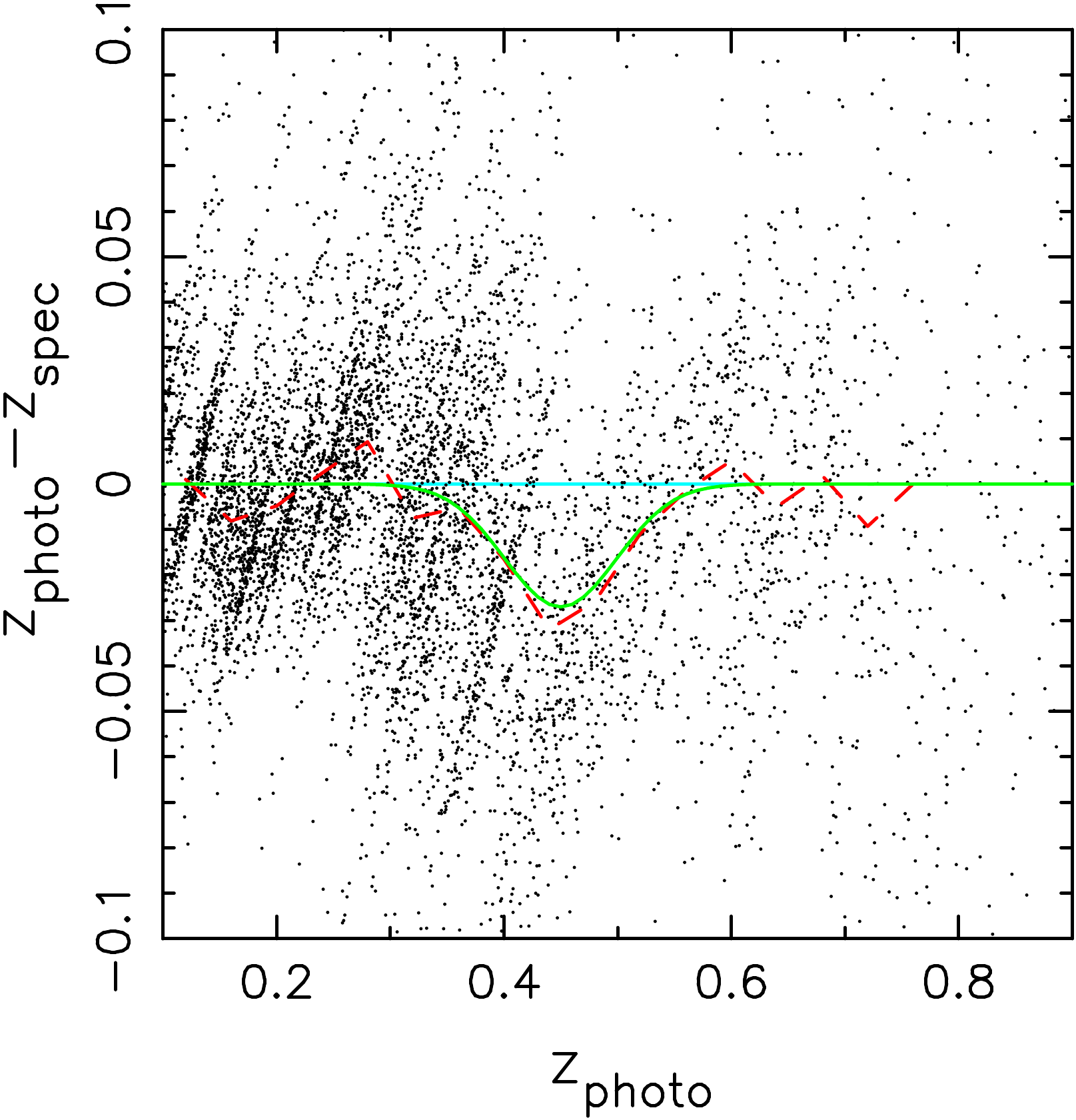}
\caption{Photometric redshift errors as a function of $z_{photo}$.  Photometric redshifts were systematically underestimating the true spectroscopic redshifts around $z=0.45$.  A Gaussian function (green solid line) was fit to the median redshift residual (red dashed line), and used to correct the photometric redshifts.  The banding is due to large scale structure.}
\label{fig:zcorr}
\end{center}
\end{figure}

Photometric redshifts were determined from imaging in the $B_W, R$ and $I$ optical bands and the 3.6, 4.5, 5.4 and 8.0$~\mu$m infrared bands, using the ANNz empirical photometric redshift code \citep{firth03, colli04}, as described in \citet{brown08}.  ANNz systematically underestimated redshifts by about $5\%$ around $z=0.45$.  A Gaussian function was fit to the median of the redshift residual, then used to correct the photometric redshifts, as shown in Figure \ref{fig:zcorr}.  The redshift correction used was
\begin{equation}
z_p = z_p' + 0.027e^{-205(z_p'-0.45)^2}
\end{equation}
where $z_p'$ is the uncorrected photometric redshift and $z_p$ is the corrected photometric redshift.  Using the uncorrected redshifts changed the measured spatial clustering of the affected samples by less than $7\%$ and had no effect on our conclusions.

\begin{figure}
\begin{center}
\includegraphics[width=\columnwidth]{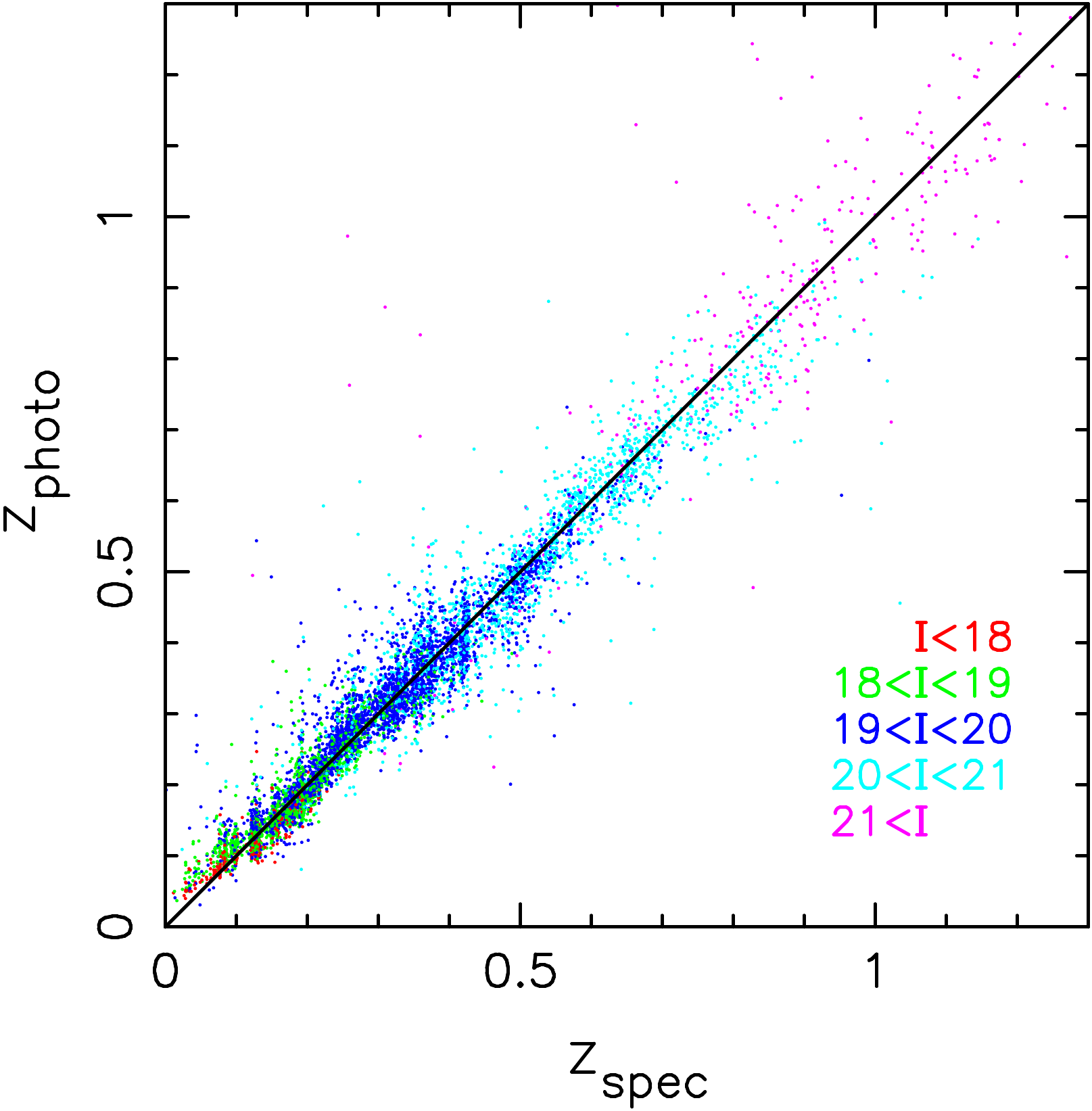}
\caption{A comparison of the corrected photometric redshifts and spectroscopic redshifts for the $24~\mu$m sources in our sample.  Data points are coloured by $I$-band magnitude.  The typical uncertainty in the photometric redshifts is $\sigma_{\Delta z/(1+z)} \simeq 0.03$, with a catastrophic failure rate of $1.6\%$ at $z\simeq0.3$ and increasing to $6.2\%$ at $z\simeq0.9$.}
\label{fig:zpvszs}
\end{center}
\end{figure}

\begin{figure*}
\begin{center}
\resizebox{2\columnwidth}{!}{\plottwo{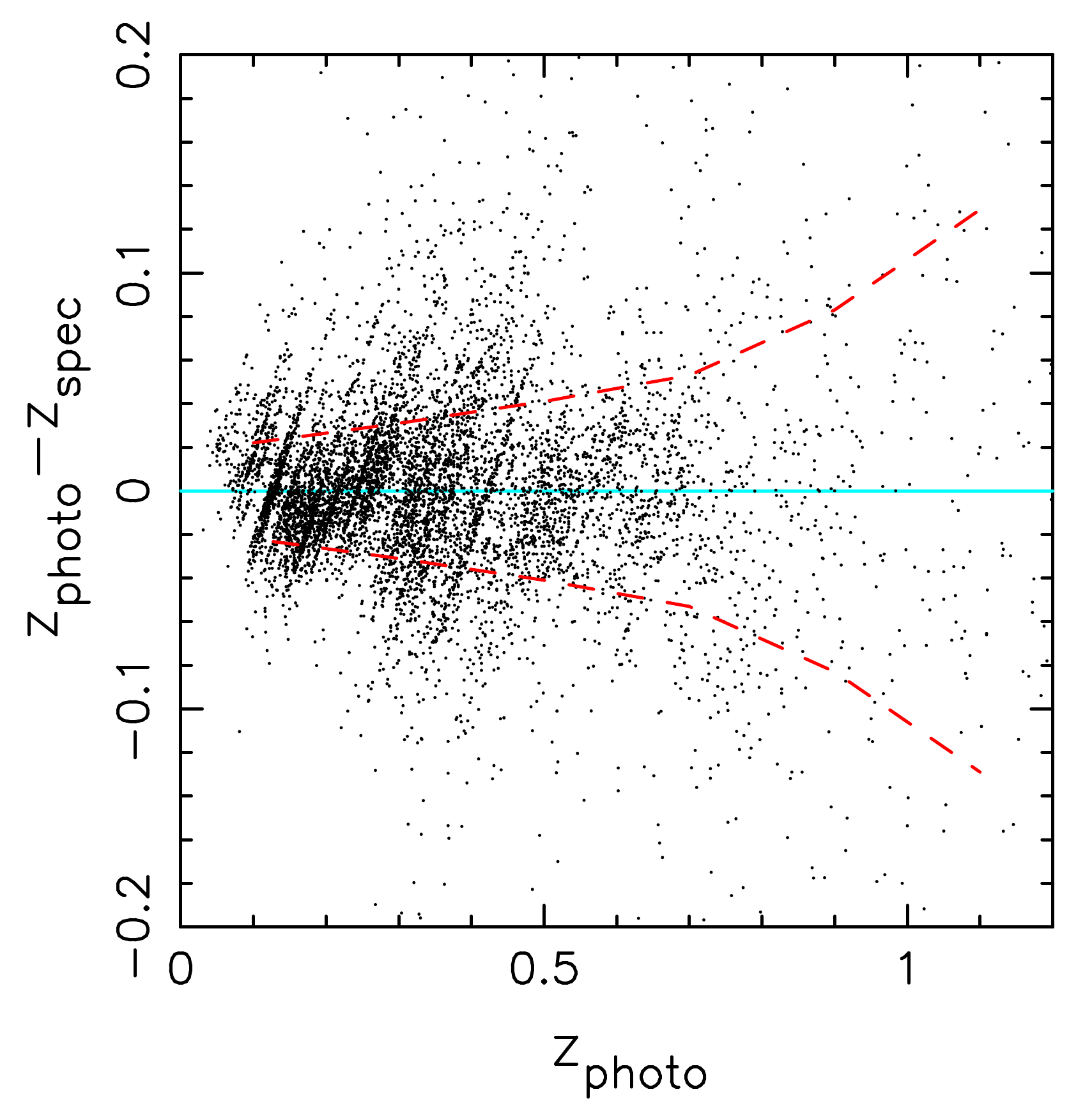}{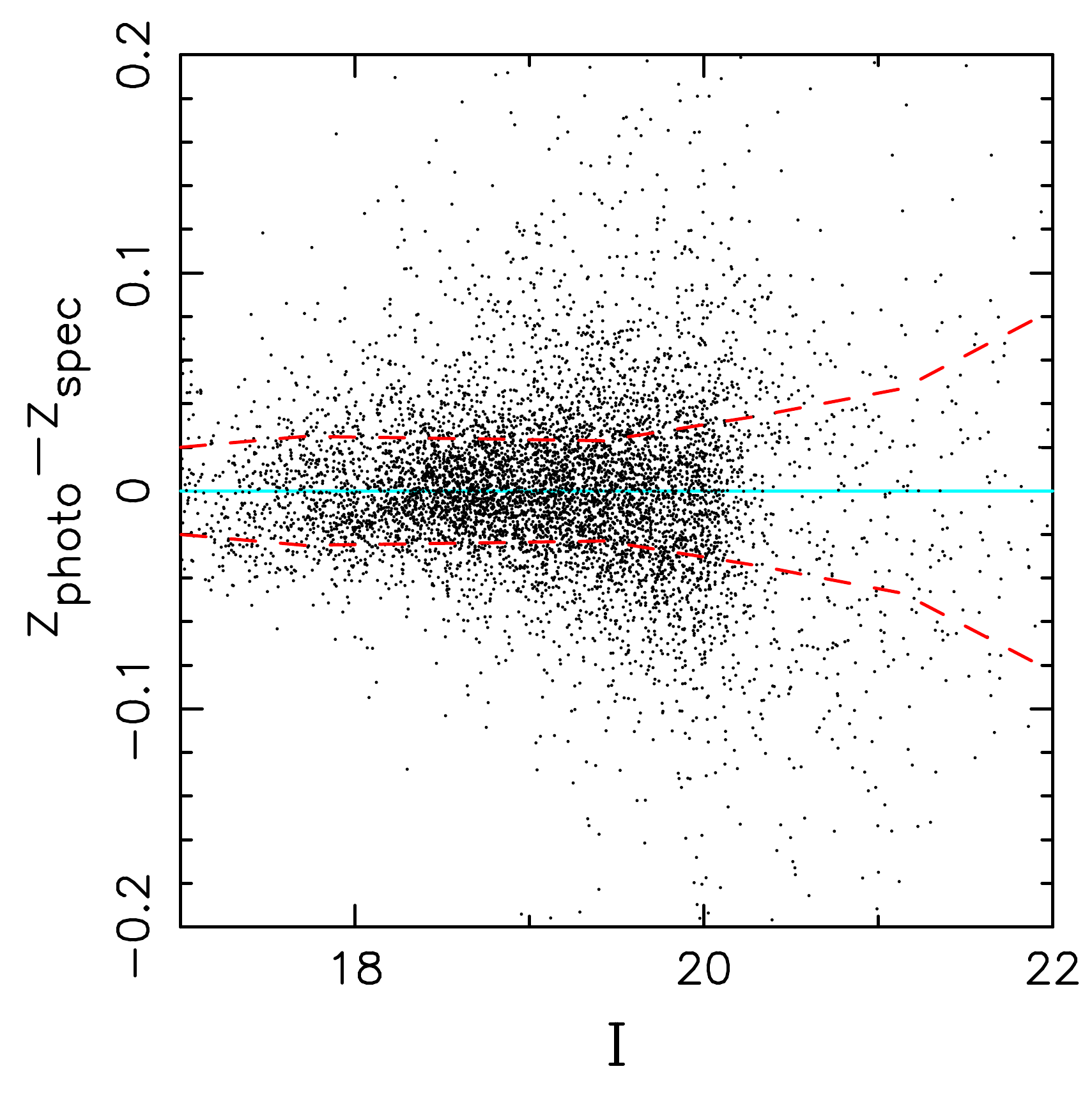}}
\caption{The redshift residual as a function of photometric redshift (left) and $I$-band magnitude (right) for the $24~\mu$m sources in our sample.  These are used to estimate the uncertainties in our photometric redshifts, which are an increasing function of redshift and $I$-band magnitude.  The typical uncertainty in the photometric redshifts is $\sigma_{\Delta z/(1+z)} \simeq 0.03$.  The red dashed lines enclose 68\% of the data points at each redshift.  The banding is due to large scale structure.}
\label{fig:pzmsz}
\end{center}
\end{figure*}

Figure \ref{fig:zpvszs} shows a comparison between photometric and spectroscopic redshift for objects in our star forming galaxy sample, for comparison with previous work.  We also show the redshift residuals in Figure \ref{fig:pzmsz} as a function of two measurable quantities: photometric redshift and $I$-band magnitude.  From these 6813 objects the uncertainties in the photometric redshifts were estimated as a function of $I$-band magnitude and redshift to be $\sigma_{photoz} = 0.028(1.2)^{I+5z-21}$.  We find $\sigma_{\Delta z/(1+z)} \simeq 0.03$ where $\Delta z = |z_{photo} - z_{spec}|$.  We define a catastrophic failure for photometric redshifts as $\Delta z/(1+z)>3\sigma$, as used by \citet{palam13}.  At $z\simeq0.3$ this failure rate is $1.6\%$ but increases to $6.2\%$ at $z\simeq0.9$.  If we adopt a more common definition of catastrophic failures, such as $\Delta z/(1+z)>0.15$ \citep{karta10}, we obtain much smaller failure rates of $0.4\%$ at $z\simeq0.3$ increasing to $3.7\%$ at $z\simeq0.9$.

We repeated the clustering analysis in section \ref{sec:cluster} using SED template photometric redshifts \citep{brown14}.  All measured correlation lengths were consistent within $1\sigma$ uncertainties, although for $z>0.6$, $r_0$ is typically $10-15\%$ larger when using the template redshifts.  We compare both sets of photometric redshifts and find no systematic offset between them.  The uncertainties in the template-derived photometric redshifts were typically double the uncertainties in the ANNz redshifts, so we adopt the latter for all further analysis.  While it is plausible that another photometric redshift code may give better redshifts, for the two sets we tested we get consistent results and conclusions.

\subsection{Optical Colours}

\begin{figure*}
\begin{center}
\resizebox{2\columnwidth}{!}{\plottwo{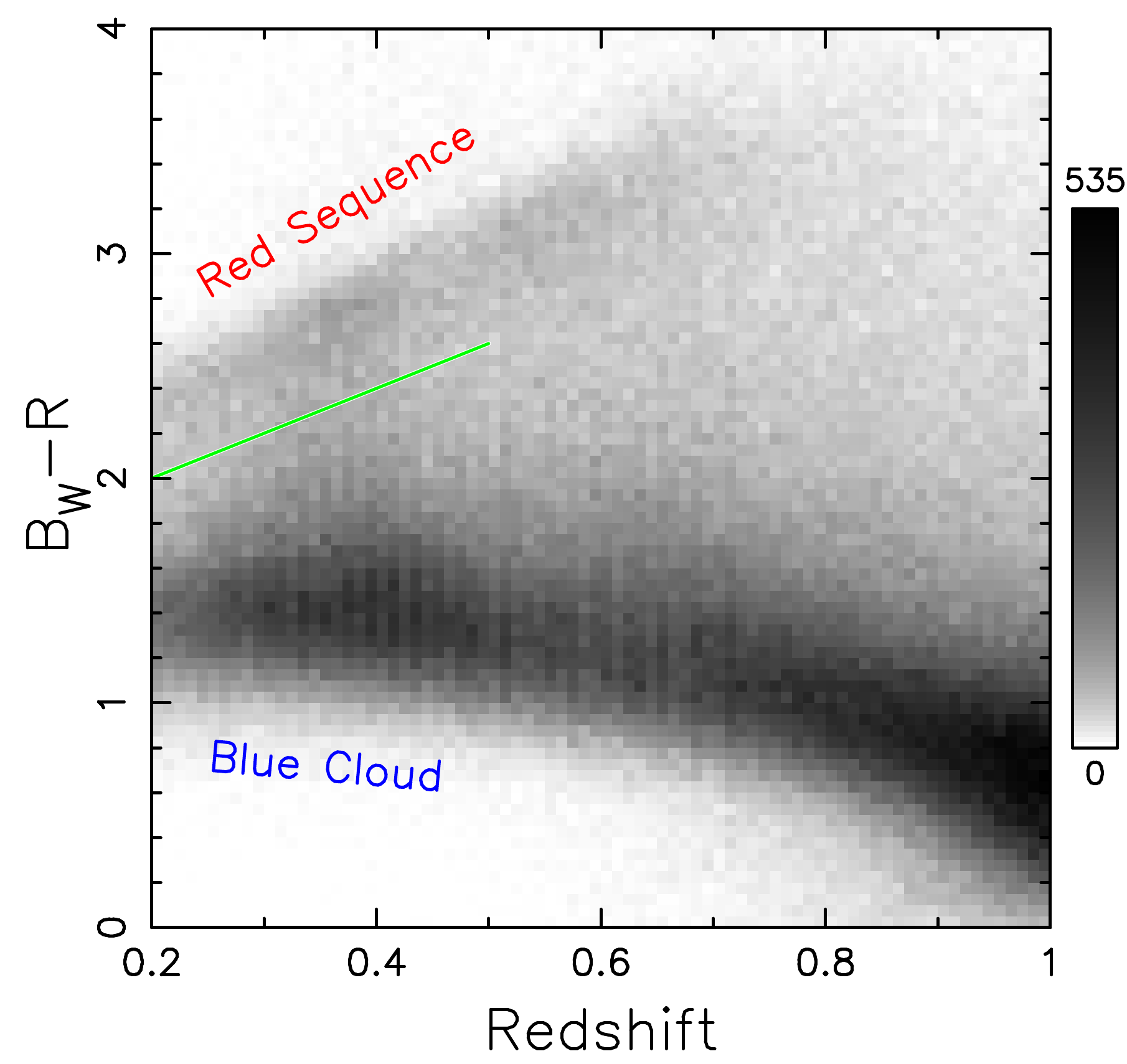}{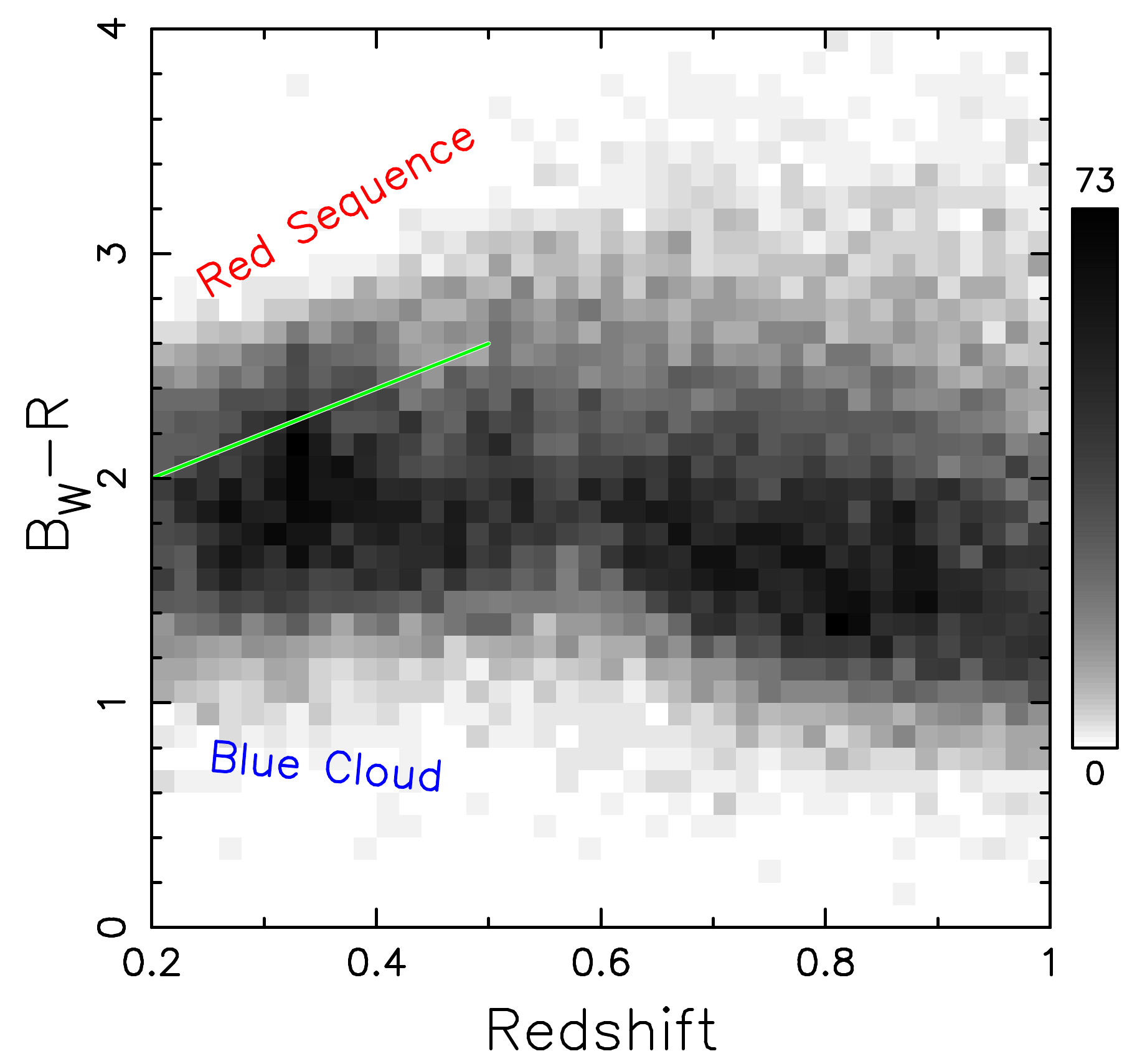}}
\caption{$B_W-R$ color as a function of photometric redshift for all galaxies in NDWFS (left) and for star forming galaxies with $F_{24~\mu\rm{m}}>0.223$~mJy (right).  The green line is a commonly used separator for blue and red galaxies out to $z\sim0.6$.  The red sequence is above the line, while the blue cloud sits below it.  The majority of our star forming sample reside within the blue cloud, but the locus is redder than that for all blue galaxies, so galaxies with the highest star formation rates are redder than typical blue galaxies.}
\label{fig:BwR}
\end{center}
\end{figure*}

Figure \ref{fig:BwR} shows the $B_W-R$ color as a function of photometric redshift for all galaxies in the NDWFS catalog and for our star forming galaxy sample.  These bands straddle the rest frame $4000$\AA~break at $z\lesssim0.5$ and are commonly used to separate red and blue galaxies.  We use the colour cut $B_W-R = 1.6 + 2z$, shown by the green line in Figure \ref{fig:BwR}.  The red sequence of passive galaxies lies above this line and the blue cloud of star forming galaxies sits below it.  The majority of galaxies in our star forming sample reside within the blue cloud, as expected for galaxies with a young stellar population, but the locus is redder than that for blue galaxies in general.  In fact, many of the galaxies are lying in the ``green valley'' between the red sequence and blue cloud.  This is in agreement with \citet{bell05}, who find that MIR galaxies with star formation contaminate the red sequence.  It has been shown that these star forming galaxies are optically red due to dust obscuration \citep[][]{weine05,willi09,bell12}.  We reproduced Figure \ref{fig:BwR} for the faintest $24~\mu$m sources in our sample and found that their locus is indeed bluer, and corresponds to that of typical blue cloud galaxies.  This is not unexpected, since \citet{weine05} and \citet{bell05} show that the optically red star forming galaxies are biased towards higher mass galaxies with higher SFRs.

\begin{figure} 
\begin{center} 
\includegraphics[width=\columnwidth]{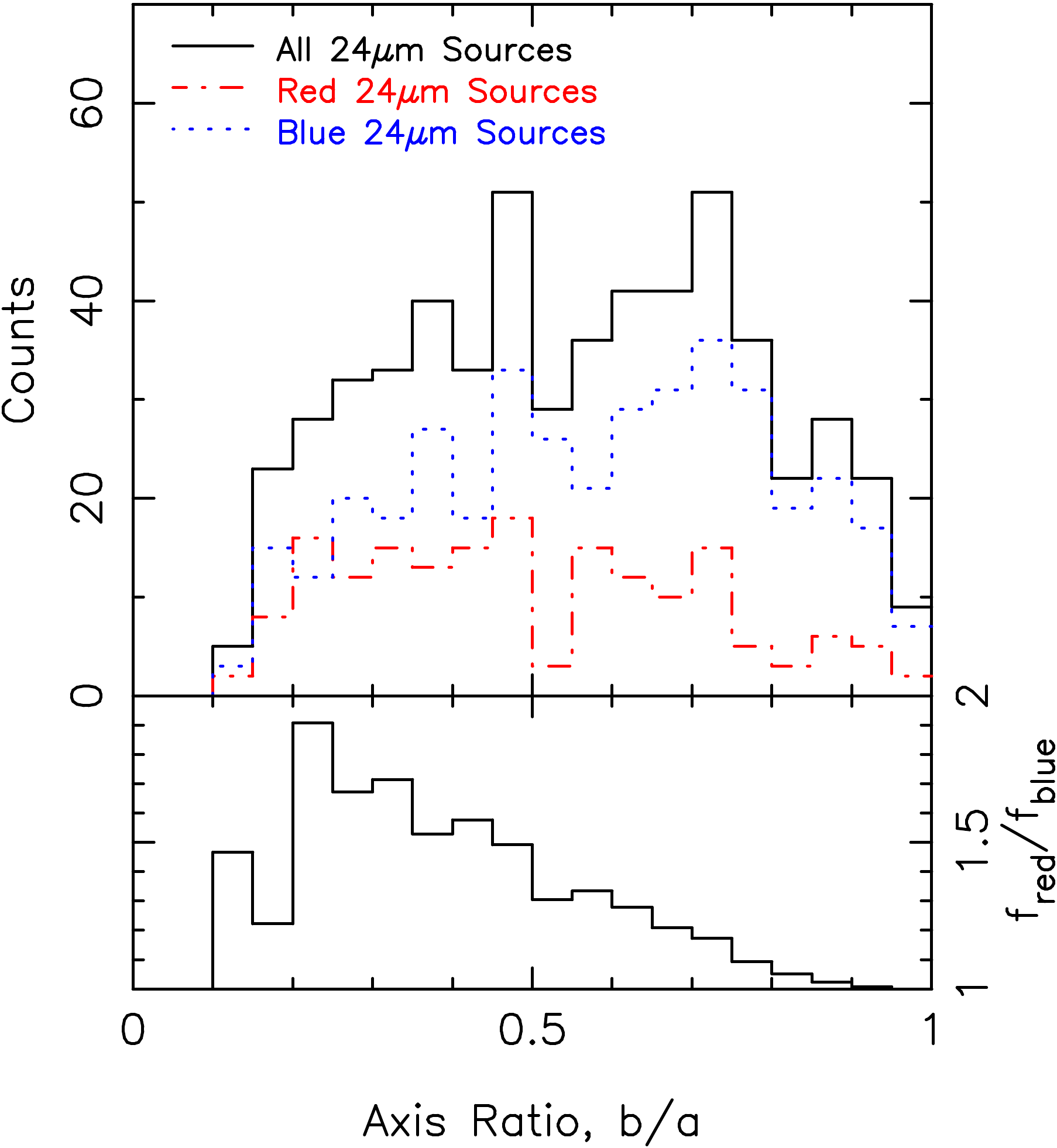}
\caption{Axis ratio distribution of star forming galaxies at $0.20<z<0.25$ based on SDSS g-band adaptive moments.  The sample is split into red and blue star forming galaxies, using the same color cut as in Figure \ref{fig:BwR}.  The bottom panel shows the ratio of the cumulative fraction of red star forming galaxies to the cumulative fraction of blue star forming galaxies.  Red star forming galaxies are $\sim1.5$ times more likely to have an axis ratio less than 0.5 than blue star forming galaxies.  A Kolmogorov-Smirnov test excludes the likelihood of the red star forming galaxies and blue star forming galaxies being drawn from the same axis ratio distribution at the $0.03\%$ significance level.  This suggests that the redder optical colors of these galaxies are due to dust, since at inclined orientations dust obscuration in disk galaxies is far greater.}
\label{fig:axis}
\end{center}
\end{figure}

The optically red galaxies in our sample must be dust obscured, since we have removed AGN.  If these optically red galaxies are dust reddened disk galaxies, then a large fraction of the sample should be roughly edge on.  As a cross check, we measured the axis ratio distributions for $0.20<z<0.25$ galaxies in our star forming sample using SDSS $g$-band adaptive moments \citep{stoug02,abaza09}.  We correct the moments for the effects of the point spread function at the position of each galaxy as described in \citet{berns02}.  This results in a small bias when objects are not well resolved \citep{hirat03}, so we only use objects when the sum of the adaptive second moments in the CCD row and column directions, is greater than 3 times that of the point spread function.  Figure \ref{fig:axis} shows the axis ratio histogram for all star forming galaxies, and for red and blue star forming galaxies, separated by the color cut shown in Figure \ref{fig:BwR}.  The flat distribution matches that expected for disk galaxies \citep{lamba92,ryden04}, but 57\% (27\%) of the red star forming galaxies have an axis ratio less than 0.50 (0.33), compared to only 38\% (16\%) of the blue star forming galaxies.  A Kolmogorov-Smirnov test rejects the null hypothesis that red star forming galaxies and blue star forming galaxies are selected from the same axis ratio distribution at the $0.03\%$ level.  This suggests that the redder optical colors of these galaxies are due to dust, since at inclined orientations dust obscuration in disk galaxies is far greater.

\subsection{Infrared Luminosities}

Using a fixed $24~\mu\rm{m}$ flux density limit means the minimum IR luminosity we observe differs by more than an order of magnitude between our lowest and highest redshift samples, so differences in the measured correlation lengths and halo masses may be a consequence of the differing luminosities of the samples, rather than star forming galaxies residing within different environments at different epochs.  To determine if such a luminosity dependence exists in our clustering measurements, we also measure the clustering of total infrared luminosity ($L_{TIR}$) selected samples.

\begin{figure}
\begin{center}
\includegraphics[width=\columnwidth]{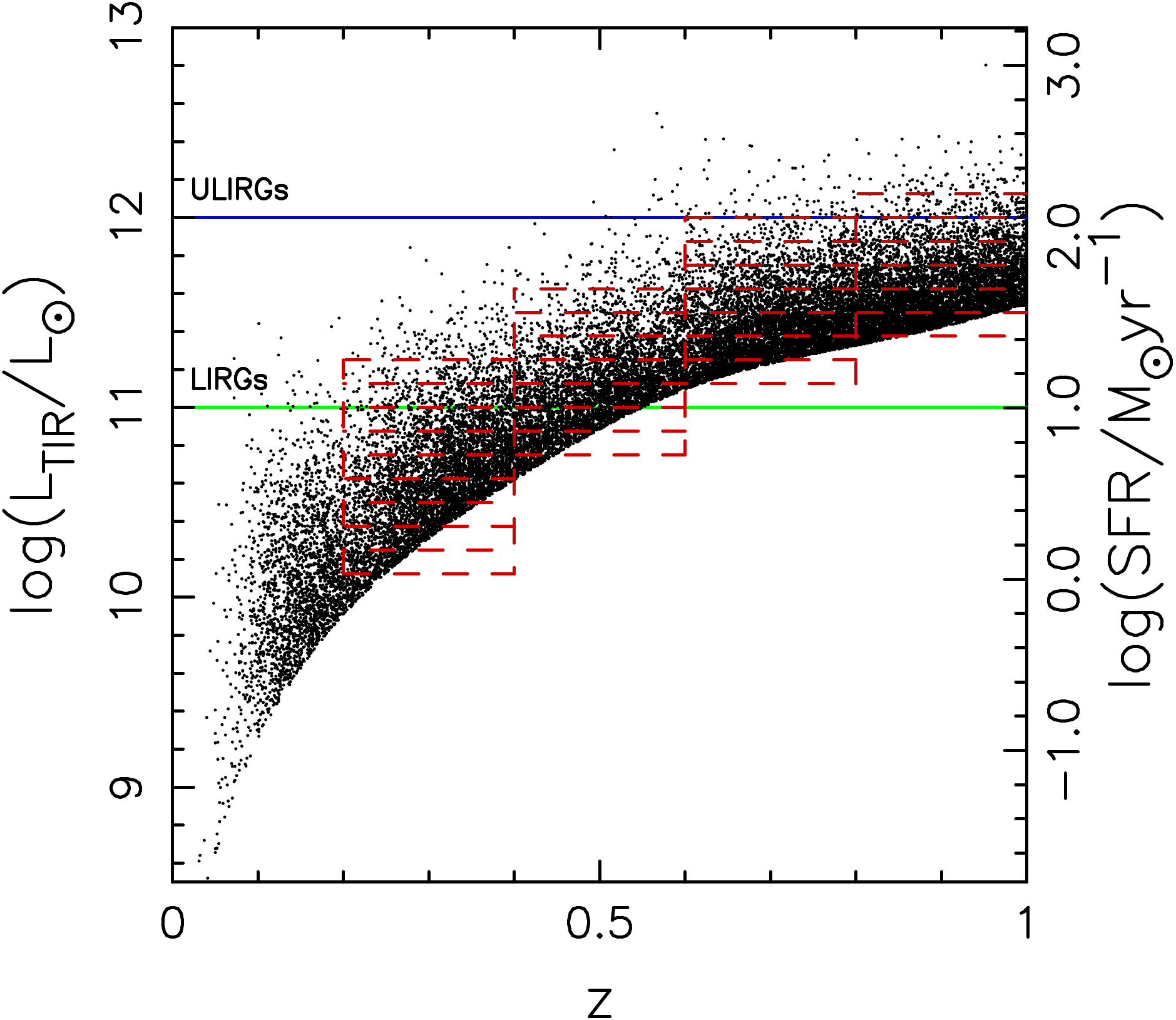}
\caption{The distribution of total infrared luminosities, for $F_{24\mu\rm{m}}>0.223~$mJy.  The green and blue lines show the luminosity thresholds for LIRGs and ULIRGs.  The red dashed lines show the overlapping $0.25~\rm{dex}$ $L_{TIR}$ samples used.}
\label{fig:LTIR}
\end{center}
\end{figure}

We estimated SFRs and $L_{TIR}$ from the $24~\mu\rm{m}$ flux densities, using the power-law fits to the SFR-$F_{24\mu\rm{m}}$ relation provided by \citet{rieke09}, but we incorporate the modifications of \citet{rujop13}, which correct for the systematic over estimation of $L_{TIR}$ at high redshift.  They define $L_{TIR}$ as the luminosity obtained by integrating the SED from $5~\mu$m to $1000~\mu$m.  Figure \ref{fig:LTIR} shows the $L_{TIR}$ and SFR distribution for our star forming galaxies.  We split the data into overlapping $L_{TIR}$ selected samples with bin widths of $0.25~\rm{dex}$, discarding any sample with $\lesssim400$ objects because we cannot reliably estimate correlation functions with such low pair counts.  The typical uncertainty in $z_{photo}$ at the mean redshift and $24~\mu$m flux density corresponds to an uncertainty in $L_{TIR}$ of $0.09~\rm{dex}$, which is smaller than the scatter in the $L_{24\mu\rm{m}}$--$L_{TIR}$ relation of $0.13~\rm{dex}$ found by \citet{rieke09}.

\section{Redshift Distribution}
\label{section:zdist}

\begin{figure*}
\begin{center} 
\resizebox{2\columnwidth}{!}{\plottwo{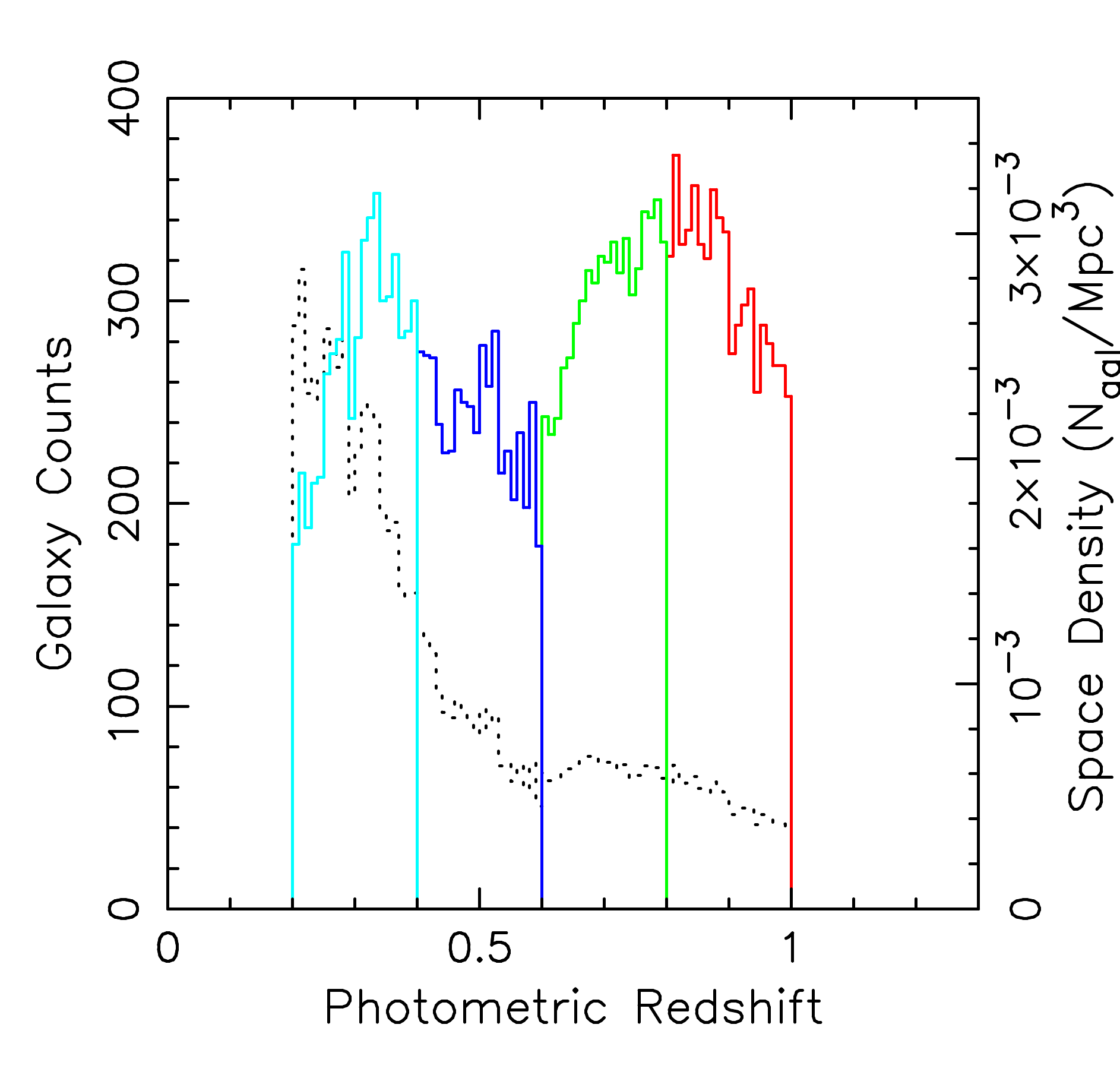}{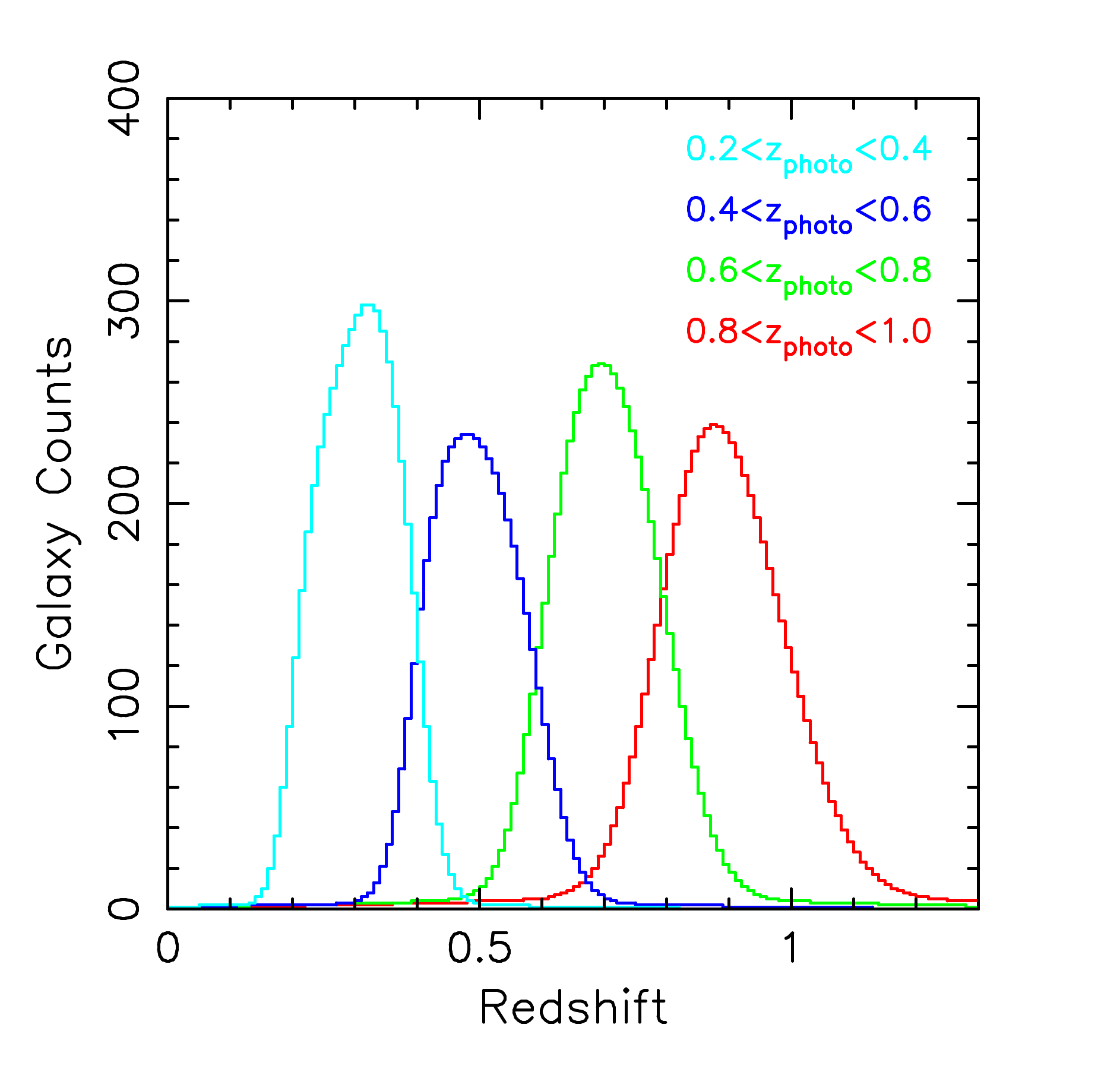}}
\caption{The photometric (left) and model (right) redshift distributions of star forming galaxies with $F_{24~\mu\rm{m}} > 0.223$~mJy.  To model the true redshift distribution, we treat each photometric redshift as Gaussian probability distribution, with its width determined by the uncertainty in the photometric redshift.  Uncertainties in photometric redshifts are modeled as an increasing function of redshift and $I$-band magnitude.  If we did not account for the uncertainties in the photometric redshifts we would underestimate the spatial clustering.  The dotted line shows the comoving space density for the same galaxy sample.  This shows that the second peak in the redshift distribution is caused by the increasing volume of redshift bins with increasing redshift.}
\label{fig:zdist}
\end{center}
\end{figure*}

Figure \ref{fig:zdist} shows the photometric redshift distribution of our star forming galaxy sample.  The distribution is shaped by four factors: the volume of each redshift bin, the galaxy luminosities visible within each redshift bin, evolution of the IR luminosity function, and spectral features entering the $24~\mu$m band.  It has been shown that the number density of luminous IR galaxies (LIRGs) increases with redshift \citep[e.g.][]{leflo05,caput07}, which increases the number of objects in our highest redshift bins.  As redshift increases, so does the comoving volume of each redshift bin, so the number of sources in each bin increases.  This is the cause of the second peak in the redshift distribution, centred at $z\approx0.85$, which is predicted by the models of \citet[][Figure A5]{lacey08} and also observed in $24~\mu$m redshift distributions by \citet{leflo05}, \citet{desai08} and \citet{magli08}.  Conversely, as redshift increases we only see increasingly more luminous objects, due to our $24~\mu$m flux limit.  As shown by the dotted line in Figure \ref{fig:zdist}, the comoving space density decreases steadily with redshift, as fainter objects fall below our flux limit.

The primary MIR spectral features that enter the $24~\mu$m bandpass, are emission from polycyclic aromatic hydrocarbons (PAH) at $11.3, 12.0, 12.7, 17, 18.9~\mu$m and Ne II at $12.8~\mu$m, which are all indicators of hot, young stars \citep[][]{smith04,desai08,treye10}.  While PAHs must contribute to the IR emission of galaxies, these spectral lines are narrow compared to the width of the MIPS $24~\mu$m bandpass, so they do not cause noticeable features in the redshift distribution because they increase the number of galaxies in many adjacent redshift bins.  The only clear feature in the redshift distribution is the drop in space density at $z \approx0.6$, which corresponds to the trough in the MIR galaxy spectrum at $14~\mu\rm{m}\lesssim\lambda\lesssim16~\mu\rm{m}$, where there is an absence of PAH emission \citep[e.g.][]{smith04,treye10}.

The main source of uncertainty in the inferred spatial clustering is due to the shape of the redshift distribution used in de-projecting the measured angular clustering.  We select our volume limited samples based on photometric redshifts.  If spectroscopic redshifts were available for all galaxies in our photometric redshift limited samples, the true redshift distributions would be somewhat broader than the photometric redshift distributions shown in Figure \ref{fig:zdist}.  This is due to the uncertainties in our photometric redshifts, which means some objects within our volume limited samples have true redshifts which lie outside the desired range.  If we did not account for the photometric redshift errors, we would underestimate the distances between the galaxies within each sample and underestimate the clustering of these galaxies.  If we naively just use the photometric redshift distributions in our clustering measurements we underestimate the correlation length of our $0.2<z<0.4$ sample by $8\%$, and our $0.8<z<1.0$ sample by $29\%$.

\begin{figure*}
\begin{center} 
\resizebox{2\columnwidth}{!}{\plottwo{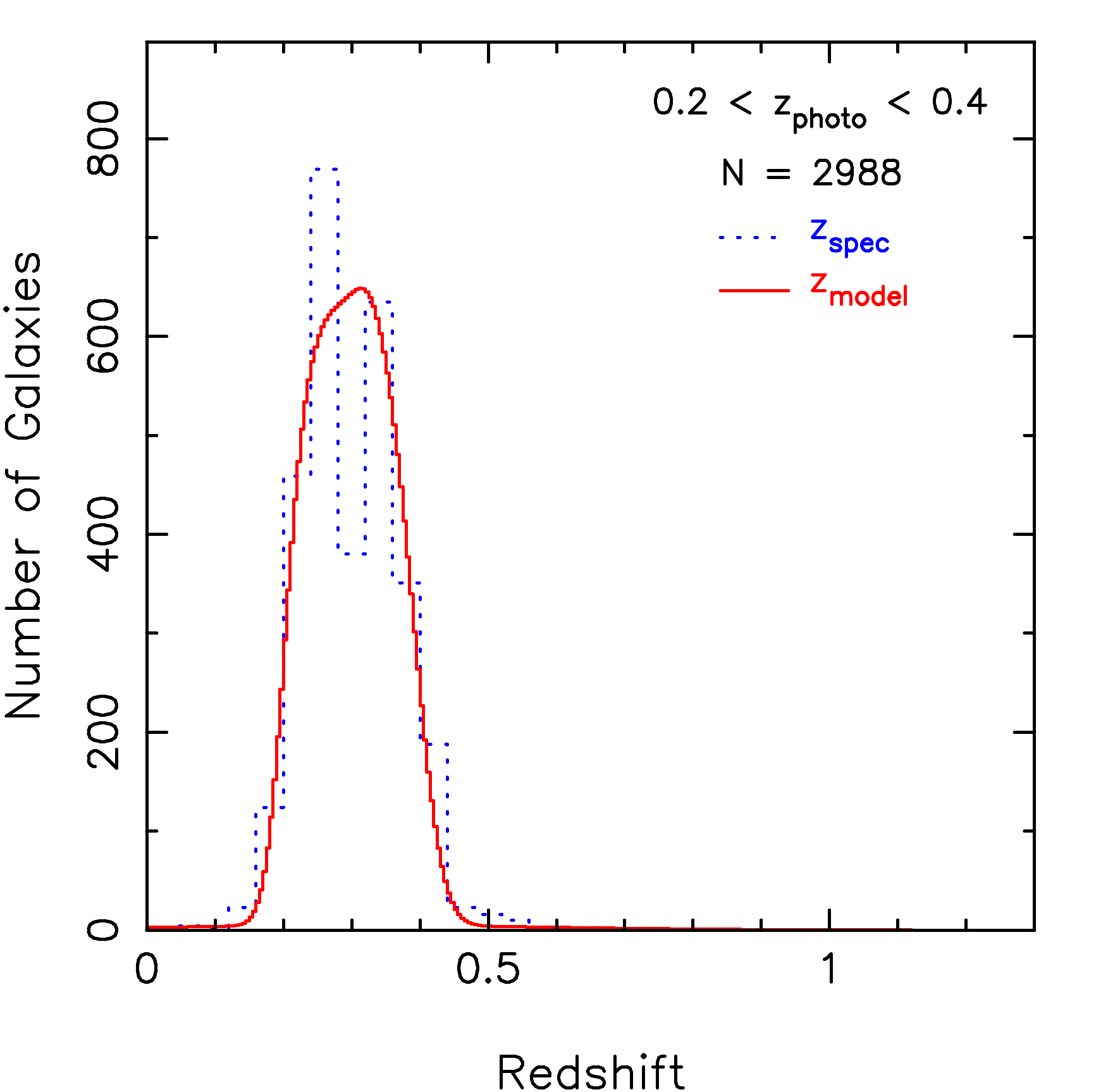}{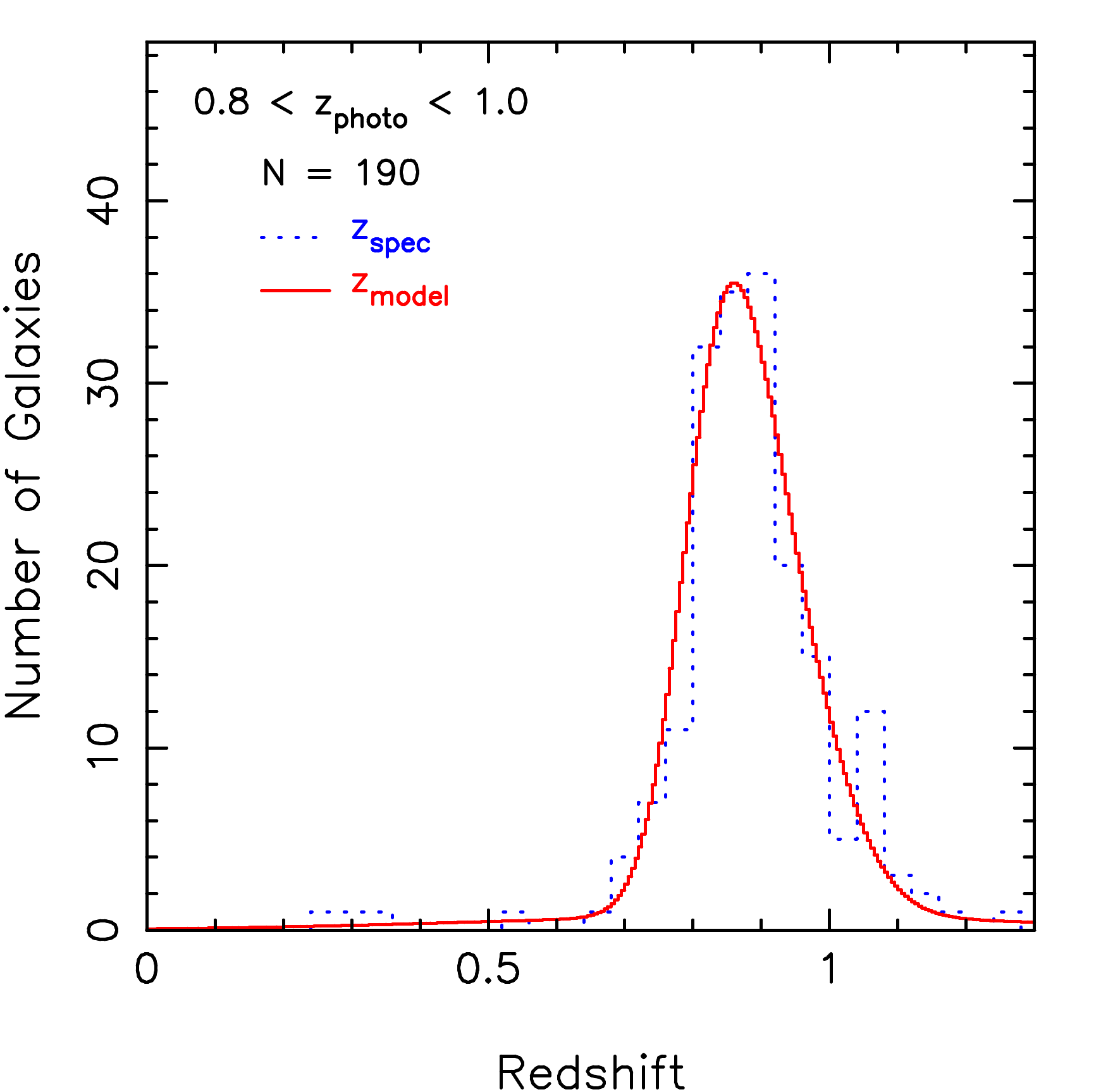}}
\caption{Spectroscopic and model redshift distributions for objects in our lowest and highest photometric redshift selected samples.  The model redshift distribution is calculated from the photometric redshifts.  The agreement of the two distributions shows that our model for correcting the photometric redshift distribution works well.  The model redshift distributions are somewhat broader than the photometric redshift distributions, due to uncertainties in the photometric redshifts.  If we did not correct our photometric redshift distributions in this way, we would underestimate the clustering of each sample.}
\label{fig:specandmodel}
\end{center}
\end{figure*}

To correct for the uncertainties in our photometric redshifts, the true redshift distribution was modeled by treating each photometric redshift as a probability distribution rather than a precise value, using a method similar to that of \citet{palam13}.  We model each photometric redshift $z_{photo}$ as a linear combination of two Gaussian functions centred at $z_{photo}$.  The shape of the first Gaussian function is determined by the uncertainty in the photometric redshift as described in Section \ref{sec:zphot}.  The second, broader Gaussian has a width given by $\sigma_{cata}=0.215(1+z)$ which models the photometric redshift catastrophic failures.  The contribution of each of these Gaussian functions to the probability distribution for an individual galaxy is determined by the catastrophic failure rate at $z_{photo}$.  The sum of these probability distributions for each galaxy gives a model of the true redshift distribution (Figure \ref{fig:zdist}) for galaxies in our photometric redshift limited samples.  The broadening of the redshift distribution caused by catastrophic failures only results in a small change in the final correlation length measurements, changing them by less than $0.5\sigma$ in the worst cases, and typically by less than $0.2\sigma$.  In Figure \ref{fig:specandmodel}, we select galaxies which have spectroscopic redshifts from our photometric redshift selected samples and then compare our model for the true redshift distribution to the actual spectroscopic redshift distribution. The two distributions agree well, indicating that our model for the true redshift distribution is reliable.

For the purpose of de-projecting the angular clustering to find the spatial clustering, we want our redshift distribution to appear as it would if we sampled objects over the entire sky.  With a limited sample volume, large scale structure can show up in the redshift distribution.  While photometric redshifts generally smooth out large scale structure, they can also cause some banding in the redshift distribution.  The convolved redshift distributions in Figure \ref{fig:zdist} smooth out any artifacts and allows us to mimic the spectroscopic redshift distribution we would have without large scale structure.

\section{Clustering Measurements}
\label{sec:cluster}

The two-point angular correlation function $\omega(\theta)$ gives the excess probability of finding two galaxies separated by an angle $\theta$ on the celestial sphere with respect to a random distribution.  Similarly the two-point spatial correlation function $\xi(r)$ is the excess probability of finding two galaxies separated by a distance $r$, with respect to a random distribution.  To estimate the angular correlation function, our star forming galaxies were compared to random distributions of galaxies.  We sampled the random catalogs from the same region mask applied to the observed catalog (Figure \ref{fig:gals}), so as not to bias the measured clustering.  We measured $\omega(\theta)$ using the \citet{landy93} estimator
\begin{equation}
\label{eqn:cest2}
\hat{\omega}(\theta)=\frac{DD-2DR+RR}{RR}
\end{equation}
where $DD$, $DR$ and $RR$ are the number of galaxy-galaxy, galaxy-random and random-random pairs at each angular separation.  We also measured $\omega(\theta)$ using the \citet{hamil93} estimator and obtained almost identical results.  Pair counts from 20 random galaxy sets were averaged to reduced random Poisson noise.

These estimators of the correlation function are subject to the integral constraint \citep[][]{groth77},
\begin{equation}
\iint\hat{\omega}(\theta)d\Omega_1 d\Omega_2\simeq 0
\end{equation}
where $\theta$ is the angle between solid angle elements $d\Omega_1$ and $d\Omega_2$.  This states that the measured correlation function summed over the whole field will be approximately zero, which results in a systematic underestimate of the actual clustering for small fields.  If the number density fluctuations in the volume are small, then this can be approximately corrected by adding a constant value to the measured angular correlation function, $\omega(\theta)=\hat{\omega}(\theta)+\omega_\Omega$, where
\begin{equation}
\label{eqn:intconst}
\omega_\Omega=\frac{1}{\Omega^2}\iint\hat{\omega}(\theta)d\Omega_1 d\Omega_2
\end{equation}
and $\Omega$ is the area of the field.  For our imaging of the Bo\"otes field we found this correction to be
\begin{equation}
\label{eqn:intconstbootes}
\omega_\Omega\approx0.6632\times10^{1.5766(1-\gamma)}\times\omega(1')
\end{equation}
which is typically $2.5\%$ of $\omega(1')$.

Poisson random errors in the number of pair counts are commonly used for estimation of uncertainties in the measured correlation function.  These are gross underestimates, because the effective Poisson random noise is dominated by fluctuations in the number of ``structures'', not the number of galaxies.  Poisson error bars would barely be visible on our correlation functions.  Our uncertainties were estimated using an analytic approximation of the full covariance matrix, as outlined by \citet{brown08}.  This is based on the method of \citet{eisen01}, but corrects for the underestimate of uncertainties when $\omega(\theta)\gtrsim1$.  \citet[][Appendix A]{brown08} estimate the covariance matrix for their correlation functions in the same field as this paper, using both the analytic approximation, jackknife subsamples and mock catalogs.  Their comparison shows that the diagonal elements are almost identical and the off-diagonal elements are in reasonable agreement.

It has be shown empirically that the angular correlation function can be approximated by a power-law \citep[e.g.][]{groth77, norbe01} of the form
\begin{equation}
\label{eqn:poww}
\omega(\theta)=\omega(1')\left(\frac{\theta}{1'}\right)^{1-\gamma}
\end{equation}
where $\omega(1')$ is the clustering amplitude at an angular separation of 1 arcminute, where it is far less sensitive to changes in the power-law index $\gamma$.  The same index $\gamma$ can also be used to approximate the spatial correlation function with a power-law.
\begin{equation}
\label{eqn:powxi}
\xi(r)=\left({\frac{r}{r_0}}\right)^{-\gamma}
\end{equation}
where $r_0$ is the spatial scale over which there is twice the probability of finding two galaxies relative to a random distribution (i.e.\ $\xi(r_0)=1$).

\begin{figure*}
\begin{center}
\includegraphics[width=1.8\columnwidth]{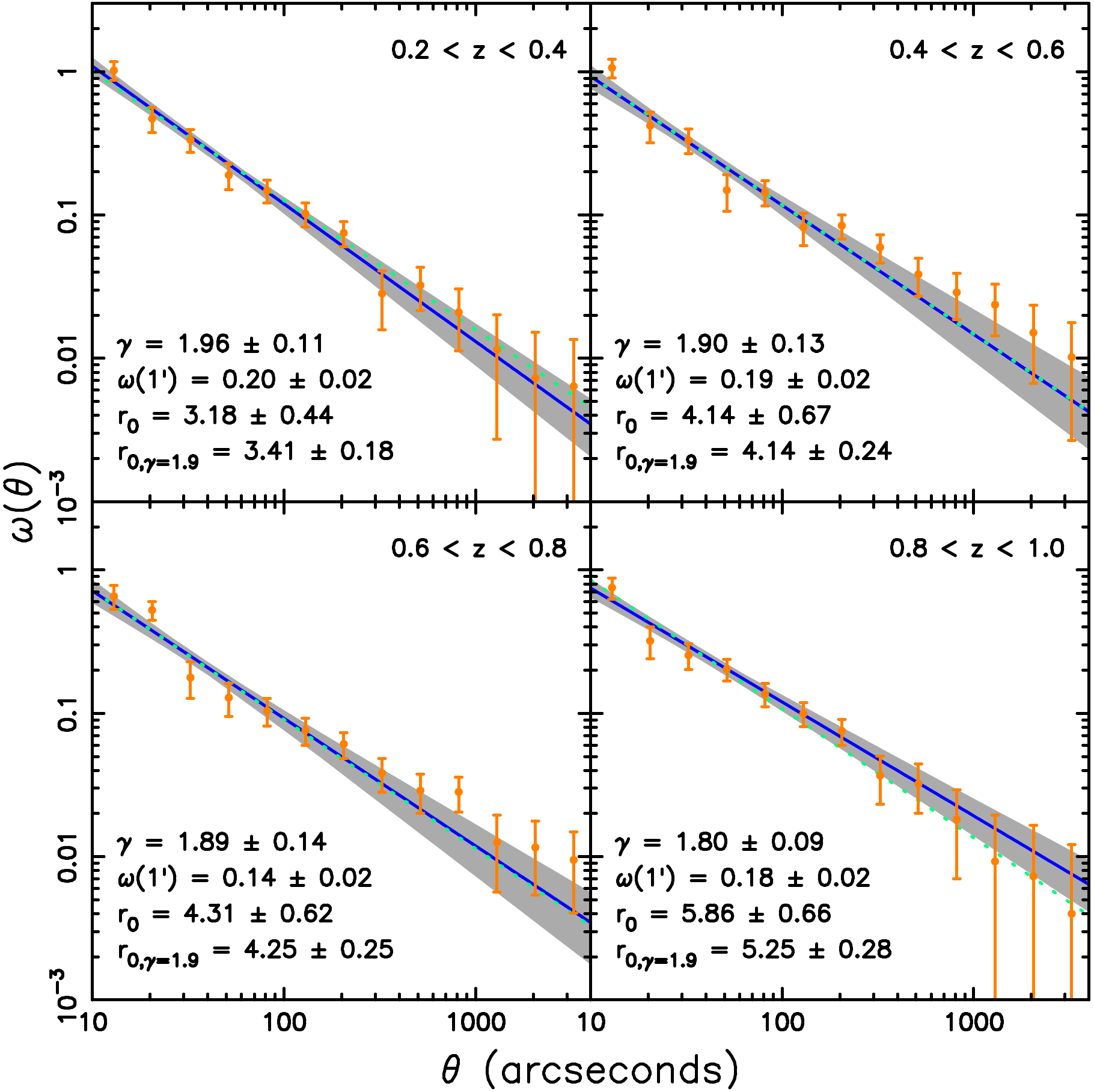}
\caption{Angular correlation functions of star forming galaxies as a function of redshift.  Power-law fits (solid lines) are obtained using the full covariance matrix, while the dotted lines are the best fit power-laws with a fixed $\gamma\equiv1.9$.  The shaded region shows power-law models within $1\sigma$ of the best fit.  The correlation length $r_0$ is in units of $h^{-1}$Mpc.}
\label{fig:correls2}
\end{center}
\end{figure*}

Power-laws fits to $\omega(\theta)$ were made including the covariance matrix in the $\chi^2$ fits, since measurements at larger separations are correlated with those at smaller separations.  Previous $24~\mu$m clustering studies used a fixed value for $\gamma$ to constrain their clustering measurements due to small sample sizes \citep{magli07,magli08,brodw08,farra06}.  Here we are able to leave both $\gamma$ and $\omega(1')$ as free parameters due to our larger sample size.  Figure \ref{fig:correls2} shows the measured angular correlation functions for each of the the $\Delta z=0.2$ redshift samples.

The spatial and angular correlation functions can be related to each other with the \citet{limbe54} equation 
\begin{equation}
\label{eqn:limber} \omega(\theta)=\frac{\int_0^\infty{}(dN/dz)\bigl\{\int_0^\infty\xi\bigl[r(\theta,z,z'),z\bigr](dN/dz')dz'\bigr\}dz\biggl.}{\bigl[\int_0^\infty{}(dN/dz)dz\bigr]^2\bigg.}
\end{equation}
where $dN/dz$ is the redshift distribution and $r(\theta,z,z')$ is the comoving distance between two objects at redshifts $z$ and $z'$ separated by an angle $\theta$.  The Limber equation integrates along two lines of sight at a given angular separation, using the spatial correlation function and the redshift distribution to determine the angular correlation function.  We used the Limber equation and the modeled redshift distributions, to find the correlation length $r_0$ which reproduces the observed angular clustering.  All of the measured clustering parameters for our $F_{24~\mu\rm{m}}$ limited samples and $L_{TIR}$ limited samples are shown in Tables \ref{table:plparams} and \ref{table:LIRparams}.

Since the fitted value of $\gamma$ affects the measured correlation length, it is useful to use a fixed value when comparing results and looking for trends in the data, so we adopt the typical value we observe, of $\gamma=1.9$.  We only show fits with $\gamma$ as a free parameter for samples with $>1000$ objects, as $\gamma$ cannot be reliably constrained with smaller samples.  Even when we fix $\gamma$, the $r_0$ values are still within $1\sigma$ of the results obtained with $\gamma$ as a free parameter.  Uncertainties shown in $r_0$ at fixed $\gamma$ assume there is no uncertainty in $\gamma$, so these are underestimates.  If we set $\gamma$ to the commonly adopted value of $1.8$, the $r_0$ values increase by approximately $0.5~h^{-1}$Mpc.

It has been shown that a power-law may not be a good fit to all galaxy correlation functions \citep[e.g.][]{zeh04,zehav11,conro06,zhe09}, because when halos contain many satellite galaxies this causes a steep increase in the small scale clustering, and can bias power-law fits and affect the measured $r_0$ and inferred halo masses.  Although a power-law does fit our data well, we do see a slight increase in the correlation function slope in our lowest angular separation bins.  If we ignore the inner two bins when fitting the power-laws we do obtain a slightly flatter typical slope, with $\gamma\approx1.83$, but the $r_0$ values obtained with $\gamma$ fixed are almost identical.

Accounting for the uncertainties in the photometric redshifts broadens the redshift distributions as discussed in Section \ref{section:zdist}.  This increases the measured correlation lengths, since the angular clustering measured is now occurring over larger distances.  Narrower redshift samples are broadened by a larger fraction of their total width, so $r_0$ is increased by a comparatively larger amount.  When a galaxy sample is split in half by redshift, we expect the $r_0$ values of narrower redshift samples to straddle that of the larger volume sample.  For example, the correlation length of our $0.6<z<1.0$ sample was $r_0=4.71\pm0.22~h^{-1}$Mpc, and when the sample was split in half we obtained $r_0=4.25\pm0.25$ and $5.25\pm0.28~h^{-1}$Mpc for the subsets, which straddle the overall value.  Similar consistency was found when all redshift samples were split in half, which is an indication that our estimation of the photometric redshift uncertainties and the modeled redshift distributions derived from these are accurate.

%
\begin{table*}
\centering
\caption{Correlation function parameters for star forming galaxy samples.}
\label{table:plparams}
\begin{tabular}{cccccccccc}
\hline \hline
Redshift & $N_{galaxy}$ & $F_{24}$ Limit & $\left<z\right>$ & $\omega(1')$ & $\gamma$ & $r_0$ & $\chi^2/dof$ & $\log(M_{halo})$ & Bias\\
 & & (mJy) & & & & $(h^{-1}$Mpc) & & $(h^{-1}M_\odot)$ & \\ \hline
$0.2<z<1.0$ & $22553$ & 0.223 & 0.62 & $0.06\pm0.01$ & $2.13\pm0.10$ & $3.34\pm0.43$ & 2.41 & $11.8^{+0.3}_{-0.3}$ & $1.08\pm0.08$ \\
$0.2<z<0.6$ & $10314$ & 0.223 & 0.40 & $0.12\pm0.01$ & $2.05\pm0.09$ & $3.24\pm0.41$ & 0.98 & $11.7^{+0.3}_{-0.4}$ & $0.95\pm0.07$ \\
$0.6<z<1.0$ & $12239$ & 0.223 & 0.80 & $0.10\pm0.01$ & $1.93\pm0.10$ & $4.60\pm0.52$ & 1.76 & $12.4^{+0.2}_{-0.2}$ & $1.42\pm0.10$ \\
$0.2<z<0.4$ & $5489$ & 0.223 & 0.31 & $0.20\pm0.02$ & $1.96\pm0.11$ & $3.18\pm0.44$ & 1.15 & $11.6^{+0.4}_{-0.6}$ & $0.90\pm0.07$ \\
$0.4<z<0.6$ & $4825$ & 0.223 & 0.50 & $0.19\pm0.02$ & $1.90\pm0.13$ & $4.14\pm0.67$ & 1.35 & $12.2^{+0.3}_{-0.4}$ & $1.15\pm0.12$ \\
$0.6<z<0.8$ & $6069$ & 0.223 & 0.71 & $0.14\pm0.02$ & $1.89\pm0.14$ & $4.31\pm0.62$ & 1.88 & $12.3^{+0.3}_{-0.3}$ & $1.31\pm0.12$ \\
$0.8<z<1.0$ & $6170$ & 0.223 & 0.89 & $0.18\pm0.02$ & $1.80\pm0.09$ & $5.86\pm0.66$ & 0.88 & $12.7^{+0.2}_{-0.2}$ & $1.72\pm0.13$ \\ \hline
$0.2<z<1.0$ & $22553$ & 0.223 & 0.62 & $0.06\pm0.01$ & $1.90$ & $4.36\pm0.18$ & 2.92 & $12.3^{+0.1}_{-0.1}$ & $1.26\pm0.04$ \\
$0.2<z<0.6$ & $10314$ & 0.223 & 0.40 & $0.13\pm0.02$ & $1.90$ & $3.91\pm0.18$ & 1.13 & $12.1^{+0.1}_{-0.2}$ & $1.06\pm0.03$ \\
$0.6<z<1.0$ & $12239$ & 0.223 & 0.80 & $0.11\pm0.01$ & $1.90$ & $4.71\pm0.22$ & 1.64 & $12.4^{+0.1}_{-0.1}$ & $1.44\pm0.05$ \\
$0.2<z<0.4$ & $5489$ & 0.223 & 0.31 & $0.20\pm0.02$ & $1.90$ & $3.41\pm0.18$ & 1.11 & $11.8^{+0.2}_{-0.2}$ & $0.94\pm0.03$ \\
$0.4<z<0.6$ & $4825$ & 0.223 & 0.50 & $0.19\pm0.02$ & $1.90$ & $4.14\pm0.24$ & 1.24 & $12.2^{+0.2}_{-0.2}$ & $1.15\pm0.04$ \\
$0.6<z<0.8$ & $6069$ & 0.223 & 0.71 & $0.14\pm0.02$ & $1.90$ & $4.25\pm0.25$ & 1.74 & $12.3^{+0.1}_{-0.1}$ & $1.29\pm0.05$ \\
$0.8<z<1.0$ & $6170$ & 0.223 & 0.89 & $0.17\pm0.02$ & $1.90$ & $5.25\pm0.28$ & 0.93 & $12.6^{+0.1}_{-0.1}$ & $1.60\pm0.06$ \\ \hline
$0.2<z<1.0$ & $7799$ & 0.400 & 0.56 & $0.08\pm0.01$ & $2.06\pm0.13$ & $3.80\pm0.60$ & 0.64 & $12.1^{+0.3}_{-0.4}$ & $1.13\pm0.11$ \\
$0.2<z<0.6$ & $4399$ & 0.400 & 0.38 & $0.15\pm0.02$ & $2.05\pm0.11$ & $3.49\pm0.56$ & 1.16 & $11.9^{+0.4}_{-0.5}$ & $0.98\pm0.09$ \\
$0.6<z<1.0$ & $3400$ & 0.400 & 0.79 & $0.14\pm0.02$ & $1.96\pm0.14$ & $5.26\pm0.77$ & 1.01 & $12.6^{+0.2}_{-0.3}$ & $1.53\pm0.15$ \\
$0.2<z<0.4$ & $2669$ & 0.400 & 0.30 & $0.26\pm0.03$ & $2.11\pm0.11$ & $3.09\pm0.45$ & 0.97 & $11.5^{+0.4}_{-0.6}$ & $0.88\pm0.08$ \\
$0.4<z<0.6$ & $1730$ & 0.400 & 0.49 & $0.22\pm0.04$ & $1.84\pm0.19$ & $4.68\pm0.97$ & 1.16 & $12.5^{+0.4}_{-0.5}$ & $1.24\pm0.17$ \\
$0.6<z<0.8$ & $1828$ & 0.400 & 0.70 & $0.19\pm0.04$ & $1.80\pm0.21$ & $5.27\pm1.06$ & 1.61 & $12.6^{+0.3}_{-0.4}$ & $1.47\pm0.20$ \\
$0.8<z<1.0$ & $1572$ & 0.400 & 0.89 & $0.25\pm0.05$ & $1.75\pm0.13$ & $7.20\pm1.18$ & 2.12 & $13.0^{+0.2}_{-0.3}$ & $1.99\pm0.24$ \\ \hline
$0.2<z<1.0$ & $7799$ & 0.400 & 0.56 & $0.08\pm0.02$ & $1.90$ & $4.57\pm0.31$ & 0.77 & $12.4^{+0.2}_{-0.2}$ & $1.26\pm0.06$ \\
$0.2<z<0.6$ & $4399$ & 0.400 & 0.38 & $0.16\pm0.02$ & $1.90$ & $4.21\pm0.28$ & 1.21 & $12.3^{+0.2}_{-0.2}$ & $1.10\pm0.05$ \\
$0.6<z<1.0$ & $3400$ & 0.400 & 0.79 & $0.15\pm0.03$ & $1.90$ & $5.57\pm0.46$ & 0.93 & $12.7^{+0.2}_{-0.2}$ & $1.59\pm0.09$ \\
$0.2<z<0.4$ & $2669$ & 0.400 & 0.30 & $0.28\pm0.04$ & $1.90$ & $4.03\pm0.27$ & 1.13 & $12.1^{+0.2}_{-0.2}$ & $1.03\pm0.05$ \\
$0.4<z<0.6$ & $1730$ & 0.400 & 0.49 & $0.22\pm0.05$ & $1.90$ & $4.43\pm0.46$ & 1.10 & $12.4^{+0.2}_{-0.3}$ & $1.20\pm0.08$ \\
$0.6<z<0.8$ & $1828$ & 0.400 & 0.70 & $0.19\pm0.04$ & $1.90$ & $4.87\pm0.54$ & 1.52 & $12.5^{+0.2}_{-0.2}$ & $1.40\pm0.10$ \\
$0.8<z<1.0$ & $1572$ & 0.400 & 0.89 & $0.24\pm0.05$ & $1.90$ & $6.18\pm0.64$ & 2.06 & $12.8^{+0.2}_{-0.2}$ & $1.78\pm0.13$ \\
\hline
\end{tabular}
\end{table*}

\begin{table*}
\centering
\caption{Correlation function parameters for $L_{TIR}$ selected samples.}
\label{table:LIRparams}
\resizebox{2\columnwidth}{!}{
\begin{tabular}{ccccccccccc}
\hline \hline
Redshift & $N_{galaxy}$ & $\log(L_{TIR}/L_\odot)$ & $\log(\left<L_{TIR}\right>/L_\odot)$ & $\left<z\right>$ & $\omega(1')$ & $\gamma$ & $r_0$ & $\chi^2/dof$ & $\log(M_{halo})$ & Bias\\
 & & & & & & & $(h^{-1}$Mpc) & & $(h^{-1}M_\odot)$ & \\ \hline
$0.2<z<0.4*$ & $838$ & $10.125<L_{TIR}<10.375$ & 10.270 & 0.25 & $0.25\pm0.08$ & $1.90$ & $2.78\pm0.46$ & 1.81 & $11.0^{+0.6}_{-1.0}$ & $0.81\pm0.13$ \\
$0.2<z<0.4*$ & $1218$ & $10.250<L_{TIR}<10.500$ & 10.396 & 0.28 & $0.19\pm0.06$ & $1.90$ & $2.78\pm0.42$ & 0.94 & $11.0^{+0.5}_{-0.7}$ & $0.83\pm0.12$ \\
$0.2<z<0.4*$ & $1725$ & $10.375<L_{TIR}<10.625$ & 10.516 & 0.30 & $0.26\pm0.05$ & $1.90$ & $3.63\pm0.33$ & 0.71 & $11.9^{+0.3}_{-0.3}$ & $0.97\pm0.06$ \\
$0.2<z<0.4*$ & $1931$ & $10.500<L_{TIR}<10.750$ & 10.628 & 0.32 & $0.27\pm0.05$ & $1.90$ & $3.90\pm0.33$ & 2.30 & $12.1^{+0.2}_{-0.3}$ & $1.02\pm0.06$ \\
$0.2<z<0.4$ & $1616$ & $10.625<L_{TIR}<10.875$ & 10.744 & 0.33 & $0.22\pm0.05$ & $1.90$ & $3.54\pm0.38$ & 1.35 & $11.9^{+0.3}_{-0.4}$ & $0.97\pm0.06$ \\
$0.2<z<0.4$ & $1163$ & $10.750<L_{TIR}<11.000$ & 10.867 & 0.34 & $0.37\pm0.07$ & $1.90$ & $4.52\pm0.44$ & 0.95 & $12.4^{+0.2}_{-0.2}$ & $1.12\pm0.07$ \\
$0.2<z<0.4$ & $788$ & $10.875<L_{TIR}<11.125$ & 10.985 & 0.34 & $0.55\pm0.10$ & $1.90$ & $5.58\pm0.53$ & 0.61 & $12.8^{+0.2}_{-0.2}$ & $1.29\pm0.09$ \\
$0.2<z<0.4$ & $489$ & $11.000<L_{TIR}<11.250$ & 11.107 & 0.34 & $0.51\pm0.14$ & $1.90$ & $5.42\pm0.80$ & 0.56 & $12.7^{+0.3}_{-0.3}$ & $1.26\pm0.13$ \\ \hline
$0.4<z<0.6*$ & $1410$ & $10.750<L_{TIR}<11.000$ & 10.901 & 0.46 & $0.24\pm0.05$ & $1.90$ & $4.01\pm0.45$ & 0.96 & $12.2^{+0.2}_{-0.3}$ & $1.11\pm0.08$ \\
$0.4<z<0.6*$ & $2029$ & $10.875<L_{TIR}<11.125$ & 11.016 & 0.49 & $0.28\pm0.04$ & $1.90$ & $4.86\pm0.38$ & 1.86 & $12.5^{+0.2}_{-0.2}$ & $1.27\pm0.07$ \\
$0.4<z<0.6*$ & $2023$ & $11.000<L_{TIR}<11.250$ & 11.120 & 0.51 & $0.21\pm0.04$ & $1.90$ & $4.36\pm0.42$ & 1.35 & $12.3^{+0.2}_{-0.2}$ & $1.20\pm0.07$ \\
$0.4<z<0.6$ & $1444$ & $11.125<L_{TIR}<11.375$ & 11.236 & 0.52 & $0.24\pm0.05$ & $1.90$ & $4.76\pm0.52$ & 1.33 & $12.5^{+0.2}_{-0.2}$ & $1.27\pm0.09$ \\
$0.4<z<0.6$ & $857$ & $11.250<L_{TIR}<11.500$ & 11.354 & 0.52 & $0.24\pm0.08$ & $1.90$ & $4.57\pm0.77$ & 1.06 & $12.4^{+0.3}_{-0.4}$ & $1.24\pm0.13$ \\
$0.4<z<0.6$ & $459$ & $11.375<L_{TIR}<11.625$ & 11.482 & 0.53 & $0.41\pm0.14$ & $1.90$ & $5.94\pm1.08$ & 1.64 & $12.8^{+0.3}_{-0.4}$ & $1.47\pm0.19$ \\ \hline
$0.6<z<0.8*$ & $2514$ & $11.125<L_{TIR}<11.375$ & 11.289 & 0.69 & $0.14\pm0.03$ & $1.90$ & $4.12\pm0.44$ & 0.99 & $12.2^{+0.2}_{-0.2}$ & $1.26\pm0.08$ \\
$0.6<z<0.8*$ & $3418$ & $11.250<L_{TIR}<11.500$ & 11.376 & 0.71 & $0.14\pm0.03$ & $1.90$ & $4.10\pm0.36$ & 1.54 & $12.2^{+0.2}_{-0.2}$ & $1.27\pm0.07$ \\
$0.6<z<0.8$ & $2538$ & $11.375<L_{TIR}<11.625$ & 11.486 & 0.72 & $0.14\pm0.03$ & $1.90$ & $4.20\pm0.45$ & 1.54 & $12.3^{+0.2}_{-0.2}$ & $1.29\pm0.09$ \\
$0.6<z<0.8$ & $1486$ & $11.500<L_{TIR}<11.750$ & 11.607 & 0.72 & $0.17\pm0.05$ & $1.90$ & $4.67\pm0.66$ & 1.72 & $12.4^{+0.2}_{-0.3}$ & $1.38\pm0.12$ \\
$0.6<z<0.8$ & $795$ & $11.625<L_{TIR}<11.875$ & 11.727 & 0.72 & $0.29\pm0.08$ & $1.90$ & $5.96\pm0.91$ & 1.00 & $12.8^{+0.2}_{-0.3}$ & $1.61\pm0.17$ \\
$0.6<z<0.8$ & $388$ & $11.750<L_{TIR}<12.000$ & 11.849 & 0.72 & $0.31\pm0.15$ & $1.90$ & $6.08\pm1.64$ & 0.56 & $12.8^{+0.4}_{-0.5}$ & $1.63\pm0.30$ \\ \hline
$0.8<z<1.0*$ & $3472$ & $11.375<L_{TIR}<11.625$ & 11.516 & 0.89 & $0.14\pm0.03$ & $1.90$ & $4.70\pm0.40$ & 1.56 & $12.4^{+0.2}_{-0.2}$ & $1.49\pm0.08$ \\
$0.8<z<1.0*$ & $3261$ & $11.500<L_{TIR}<11.750$ & 11.614 & 0.91 & $0.17\pm0.03$ & $1.90$ & $5.26\pm0.42$ & 0.32 & $12.6^{+0.2}_{-0.2}$ & $1.62\pm0.09$ \\
$0.8<z<1.0$ & $1937$ & $11.625<L_{TIR}<11.875$ & 11.730 & 0.91 & $0.26\pm0.04$ & $1.90$ & $6.59\pm0.56$ & 0.96 & $12.9^{+0.2}_{-0.2}$ & $1.88\pm0.12$ \\
$0.8<z<1.0$ & $996$ & $11.750<L_{TIR}<12.000$ & 11.852 & 0.91 & $0.25\pm0.07$ & $1.90$ & $6.40\pm0.93$ & 0.82 & $12.8^{+0.2}_{-0.3}$ & $1.84\pm0.19$ \\
$0.8<z<1.0$ & $501$ & $11.875<L_{TIR}<12.125$ & 11.971 & 0.91 & $0.32\pm0.12$ & $1.90$ & $7.01\pm1.47$ & 0.93 & $13.0^{+0.3}_{-0.4}$ & $1.96\pm0.30$ \\ \hline
$0.2<z<0.4*$ & $1218$ & $10.250<L_{TIR}<10.500$ & 10.396 & 0.28 & $0.18\pm0.06$ & $1.55\pm0.25$ & $3.80\pm1.11$ & 0.58 & $12.0^{+0.6}_{-1.0}$ & $0.99\pm0.18$ \\
$0.2<z<0.4*$ & $1725$ & $10.375<L_{TIR}<10.625$ & 10.516 & 0.30 & $0.27\pm0.05$ & $1.90\pm0.18$ & $3.66\pm0.76$ & 0.77 & $11.9^{+0.5}_{-0.7}$ & $0.97\pm0.12$ \\
$0.2<z<0.4*$ & $1931$ & $10.500<L_{TIR}<10.750$ & 10.628 & 0.32 & $0.24\pm0.04$ & $2.15\pm0.18$ & $2.93\pm0.57$ & 2.27 & $11.3^{+0.6}_{-0.8}$ & $0.87\pm0.14$ \\
$0.2<z<0.4$ & $1616$ & $10.625<L_{TIR}<10.875$ & 10.744 & 0.33 & $0.21\pm0.04$ & $2.05\pm0.16$ & $3.00\pm0.57$ & 1.40 & $11.4^{+0.5}_{-0.7}$ & $0.88\pm0.12$ \\
$0.2<z<0.4$ & $1163$ & $10.750<L_{TIR}<11.000$ & 10.867 & 0.34 & $0.36\pm0.06$ & $1.99\pm0.15$ & $4.05\pm0.76$ & 1.01 & $12.2^{+0.4}_{-0.5}$ & $1.05\pm0.12$ \\ \hline
$0.4<z<0.6*$ & $1410$ & $10.750<L_{TIR}<11.000$ & 10.901 & 0.46 & $0.24\pm0.05$ & $1.99\pm0.18$ & $3.68\pm0.71$ & 1.03 & $12.0^{+0.4}_{-0.6}$ & $1.06\pm0.13$ \\
$0.4<z<0.6*$ & $2029$ & $10.875<L_{TIR}<11.125$ & 11.016 & 0.49 & $0.27\pm0.04$ & $2.10\pm0.18$ & $3.93\pm0.72$ & 1.78 & $12.1^{+0.4}_{-0.5}$ & $1.11\pm0.13$ \\
$0.4<z<0.6*$ & $2023$ & $11.000<L_{TIR}<11.250$ & 11.120 & 0.51 & $0.23\pm0.05$ & $1.54\pm0.14$ & $6.49\pm1.35$ & 0.78 & $13.0^{+0.3}_{-0.4}$ & $1.55\pm0.23$ \\
$0.4<z<0.6$ & $1444$ & $11.125<L_{TIR}<11.375$ & 11.236 & 0.52 & $0.24\pm0.05$ & $1.71\pm0.18$ & $5.72\pm1.22$ & 1.29 & $12.8^{+0.3}_{-0.4}$ & $1.43\pm0.21$ \\ \hline
$0.6<z<0.8*$ & $2514$ & $11.125<L_{TIR}<11.375$ & 11.289 & 0.69 & $0.14\pm0.03$ & $1.77\pm0.22$ & $4.49\pm0.87$ & 0.96 & $12.4^{+0.3}_{-0.4}$ & $1.33\pm0.16$ \\
$0.6<z<0.8*$ & $3418$ & $11.250<L_{TIR}<11.500$ & 11.376 & 0.71 & $0.14\pm0.03$ & $1.74\pm0.17$ & $4.70\pm0.80$ & 1.39 & $12.5^{+0.3}_{-0.4}$ & $1.38\pm0.15$ \\
$0.6<z<0.8$ & $2538$ & $11.375<L_{TIR}<11.625$ & 11.486 & 0.72 & $0.14\pm0.03$ & $1.60\pm0.19$ & $5.31\pm1.03$ & 1.15 & $12.6^{+0.3}_{-0.4}$ & $1.49\pm0.19$ \\
$0.6<z<0.8$ & $1486$ & $11.500<L_{TIR}<11.750$ & 11.607 & 0.72 & $0.17\pm0.05$ & $1.70\pm0.19$ & $5.52\pm1.08$ & 1.65 & $12.7^{+0.3}_{-0.4}$ & $1.53\pm0.20$ \\ \hline
$0.8<z<1.0*$ & $3472$ & $11.375<L_{TIR}<11.625$ & 11.516 & 0.89 & $0.14\pm0.03$ & $1.79\pm0.12$ & $5.14\pm0.70$ & 1.64 & $12.5^{+0.2}_{-0.3}$ & $1.58\pm0.14$ \\
$0.8<z<1.0*$ & $3261$ & $11.500<L_{TIR}<11.750$ & 11.614 & 0.91 & $0.17\pm0.03$ & $1.78\pm0.12$ & $5.89\pm0.80$ & 0.26 & $12.7^{+0.2}_{-0.2}$ & $1.74\pm0.16$ \\
$0.8<z<1.0$ & $1937$ & $11.625<L_{TIR}<11.875$ & 11.730 & 0.91 & $0.26\pm0.04$ & $1.85\pm0.15$ & $6.91\pm1.10$ & 1.04 & $12.9^{+0.2}_{-0.3}$ & $1.94\pm0.22$ \\ \hline
\end{tabular}}
\\ * Due to the $F_{24\mu\rm{m}}$ limit, these bins are not completely sampled over the redshift and $L_{TIR}$ ranges specified, so the mean redshift $\left<z\right>$, and the mean IR luminosity $\left<L_{TIR}\right>$ should be used to characterise these samples (see Figure \ref{fig:LTIR}).\\
\end{table*}

\section{Dark Matter Halo Masses}
\label{sec:halomasses}

The measured spatial clustering of our star forming galaxies was compared with that of dark matter halos to estimate the masses of the halos they reside within.  Power spectra were produced for various halo masses at the mean redshift of each sample, assuming that each halo contains one central galaxy only.  These were produced with the semi-analytic halo occupation distribution (HOD) model of \citet{selja00} using the \citet{sheth99} dark matter mass function and halo bias, the \citet{navar96} halo profiles, the \citet{bullo01} halo concentration, and the \citet{eisen98} linear power spectrum.  There are more recent models for the HOD and other descriptions for its components, but for the large scales and halo mass range we are probing it has little effect on derived halo masses.

On small scales ($\lesssim1$ Mpc) the clustering of galaxies is determined by the number of galaxies within their host dark matter halos, but on scales larger than the size of dark matter halos ($\gtrsim1$ Mpc), the clustering of galaxies is the same as that of the dark matter halos they reside within, so we directly compared $r_0$ values to estimate the halo masses of our star forming sample.  This is an approximation, since these galaxies are most likely found within a range of halo masses, but due to the steep decrease in the dark matter mass function with increasing halo mass, $r_0$ is primarily determined by the lower mass halos they reside within.  Estimating halo masses in this way will give masses slightly higher than those produced by HOD fitting which models galaxies as residing within all halos above some threshold mass \citep[e.g.][]{zhe05,brown08,zehav11}, because the contribution from higher mass halos will increase $r_0$ value obtained for the same minimum halo mass threshold.  The resulting halo masses at the mean redshift of each sample are shown in Tables \ref{table:plparams} and \ref{table:LIRparams}.

\begin{figure*}
\begin{center}
\resizebox{2\columnwidth}{!}{\plottwo{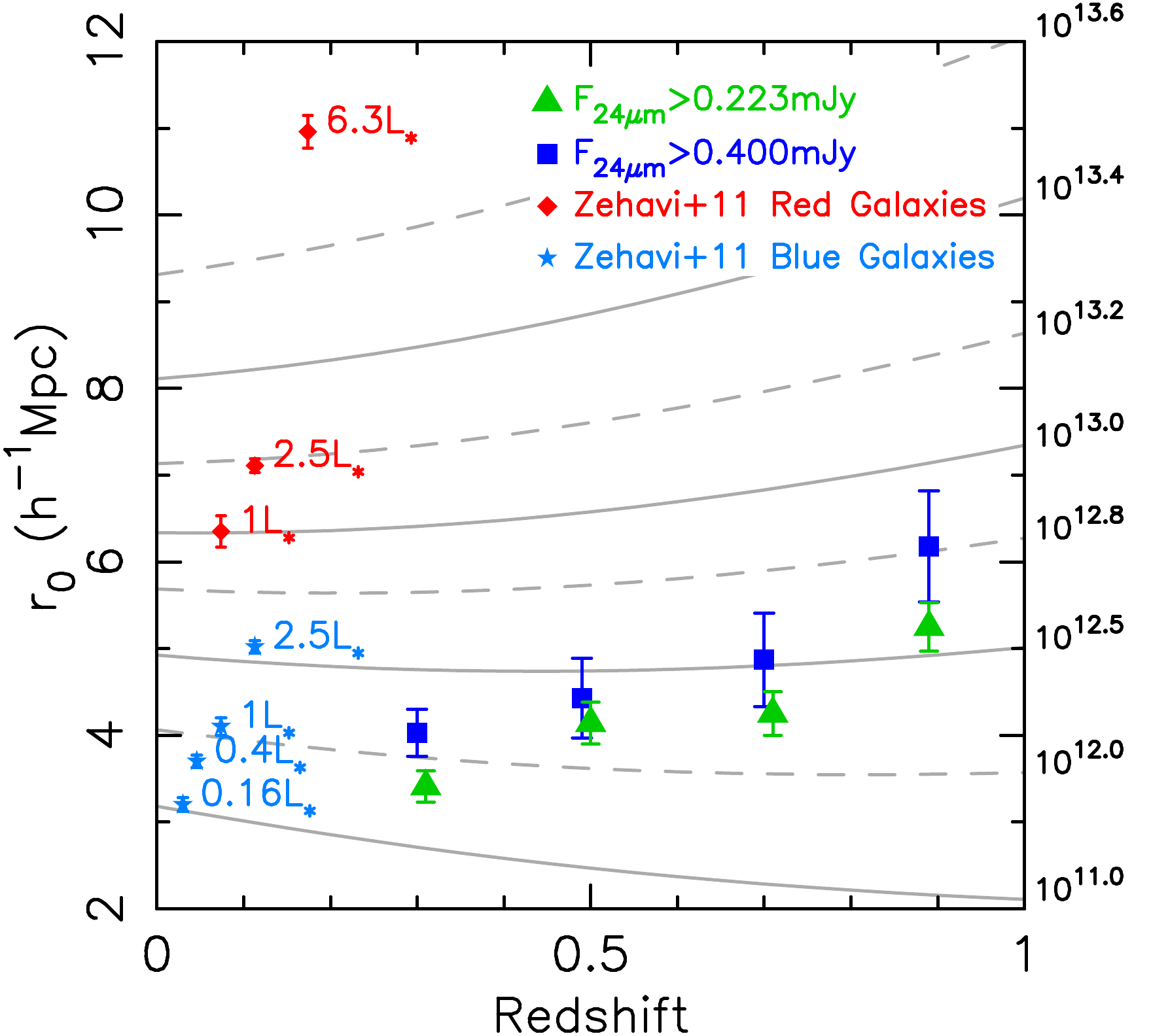}{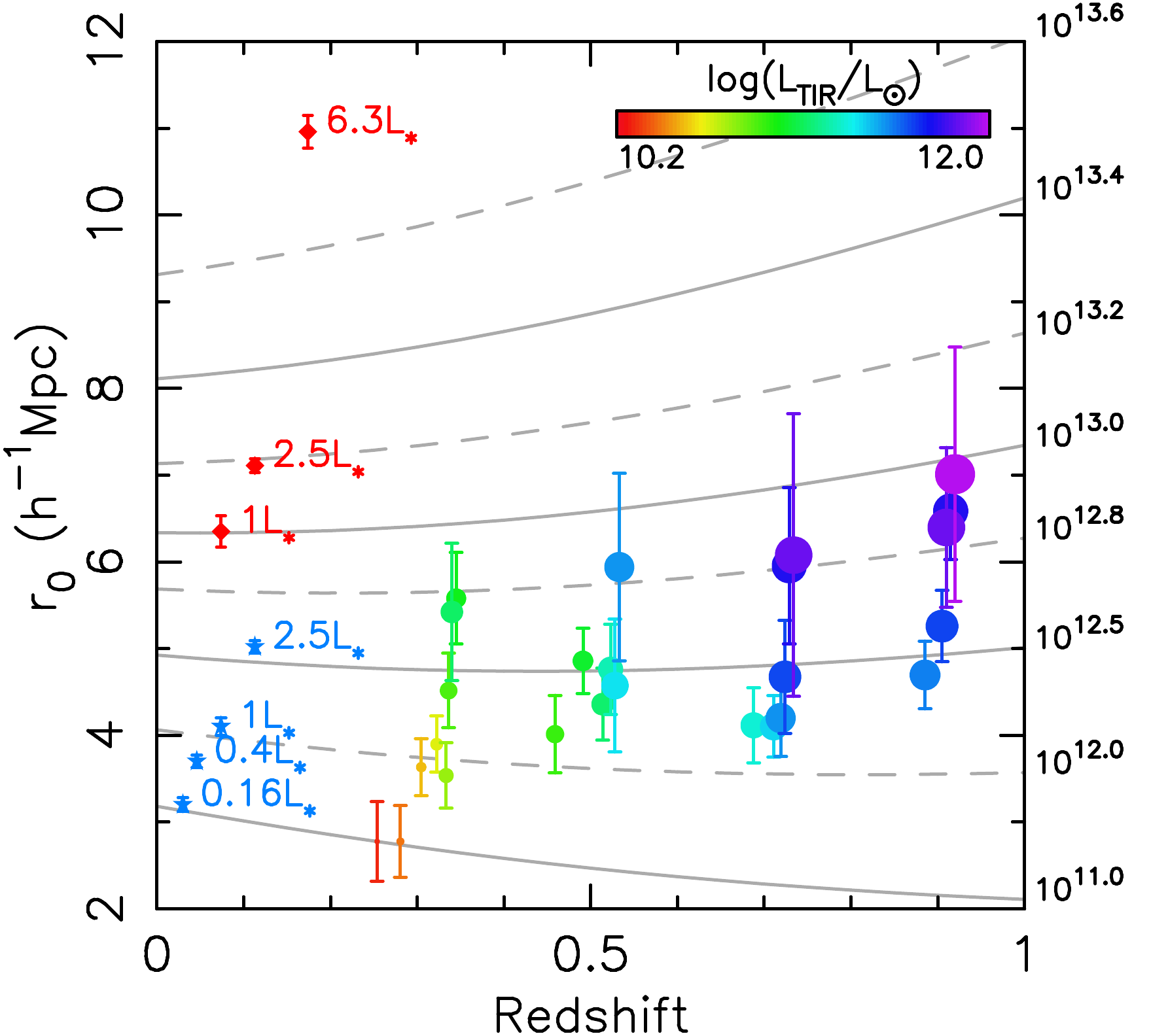}}
\caption{The clustering of star forming galaxies with a fixed $24~\mu m$ flux limit (Left) and of $L_{TIR}$ selected star forming galaxies (right).  The lines show the clustering for fixed mass dark matter halos, with masses shown in $~h^{-1}M_\odot$.  These lines are useful for comparing the typical halo masses of different galaxy samples, but they do not represent the evolution of galaxy clustering, since dark matter halos gain mass through mergers.  The size and color of data points for the $L_{TIR}$ selected samples are scaled linearly by $\log(L_{TIR})$.  Data points at low redshift show the clustering of red (diamonds) and blue (stars), luminosity selected galaxy samples from \citet{zehav11}.  Their $0.16L_*$ and $0.4L_*$ red galaxy samples have been removed for clarity, but both have $r_0\approx7~h^{-1}$Mpc.}
\label{fig:fixed_mhalo}
\end{center}
\end{figure*}

Figure \ref{fig:fixed_mhalo} shows the clustering of our star forming galaxies, and the clustering of local red and blue galaxies from \citet{zehav11}, as a function of redshift.  Overlaid  on Figure \ref{fig:fixed_mhalo} are lines showing the clustering of fixed mass dark matter halos.  These lines are useful for comparing the typical halo masses of different galaxy samples, but they do not show the evolution of galaxy clustering, since dark matter halos do not remain at a constant mass, but gain mass through mergers.  A galaxy in a $10^{12}M_\odot$ halo at $z=1$ will be in a halo $>10^{12}M_\odot$ at $z=0$.  \citet{li07} show that halos in the mass range $10^{12}-10^{14}~h^{-1}M_\odot$ typically grow in mass by ~40-60\% from $z=1$ to $0$.  All of our star forming galaxy samples have typical halo masses $\lesssim10^{13}~h^{-1}M_\odot$, which is consistent with \citet{zaude07} who find that the environment of ULIRGs is similar to that of field galaxies. 

The most massive dark matter halos gain mass through mergers, but do not move much spatially, so their correlation length remains almost constant \citep[e.g.][]{whi07}.  Lower mass halos also gain mass, but they move spatially as well, so their clustering will increase with time.  As a result, connecting them to their low redshift descendants is more difficult.  To do this, the mass assembly history of halos needs to be known from analytic models or N-body simulations \citep[e.g.][]{li07,mcbri09,fakho10,lin13}.  We integrate the analytic approximation of the mean halo mass growth rate given by \citet{fakho10} to find the mass evolution of halos from $z=1$.  This may slightly overestimate the typical growth in halo mass, since the growth rate distribution has a long positive tail.  This will cause an overestimate in the clustering evolution, however this effect is only noticeable for the most massive halos, where the correlation length of halos is sensitive to small change in halo mass.  Typically a galaxy in a halo at $z=1$ will be in the same halo at $z=0$ (i.e.~the host halo grows primarily through minor mergers), so knowing the mean evolution of dark matter halos allows us to determine the descendants of our star forming galaxy samples.  

\begin{figure*}
\begin{center}
\resizebox{2\columnwidth}{!}{\plottwo{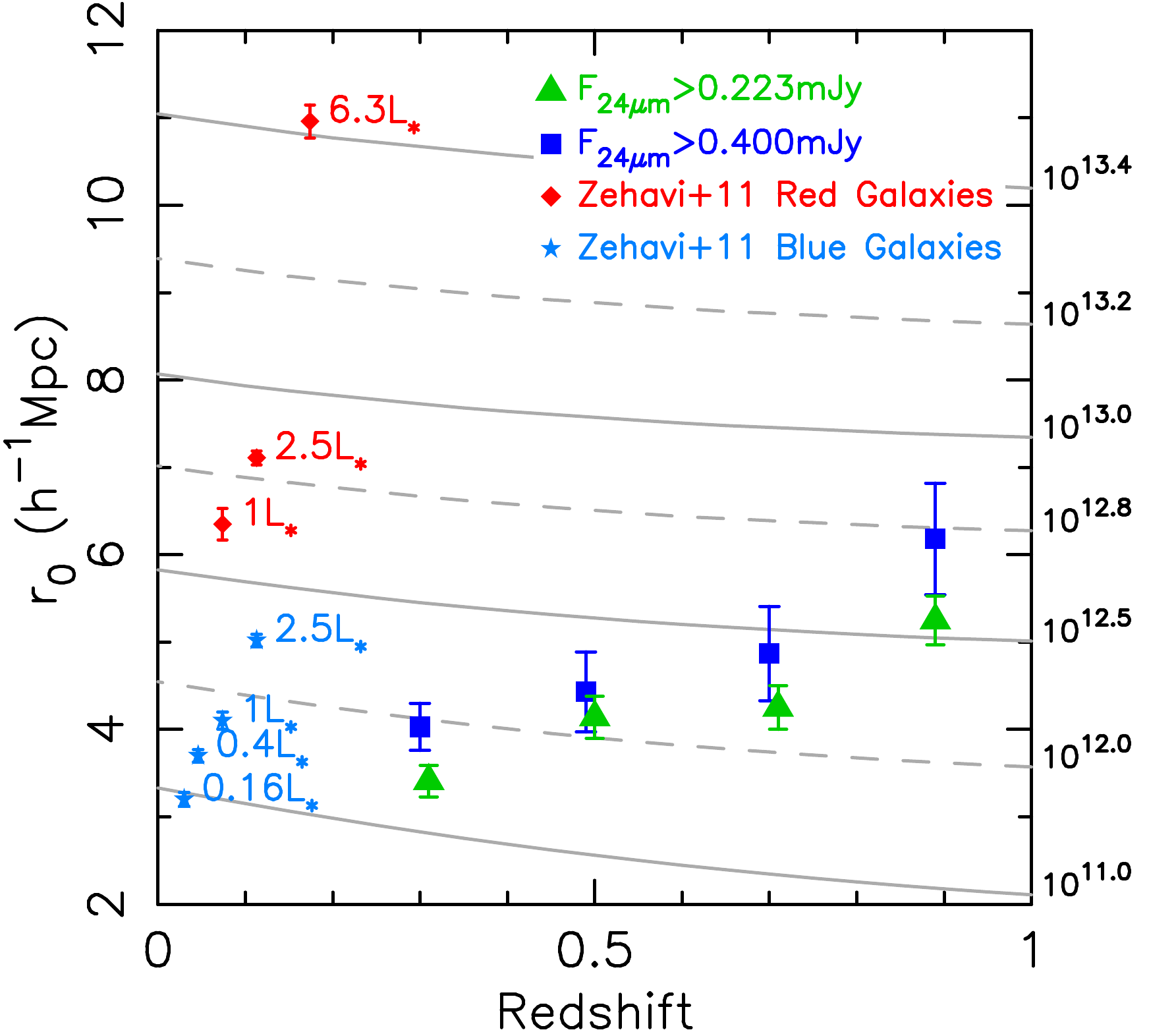}{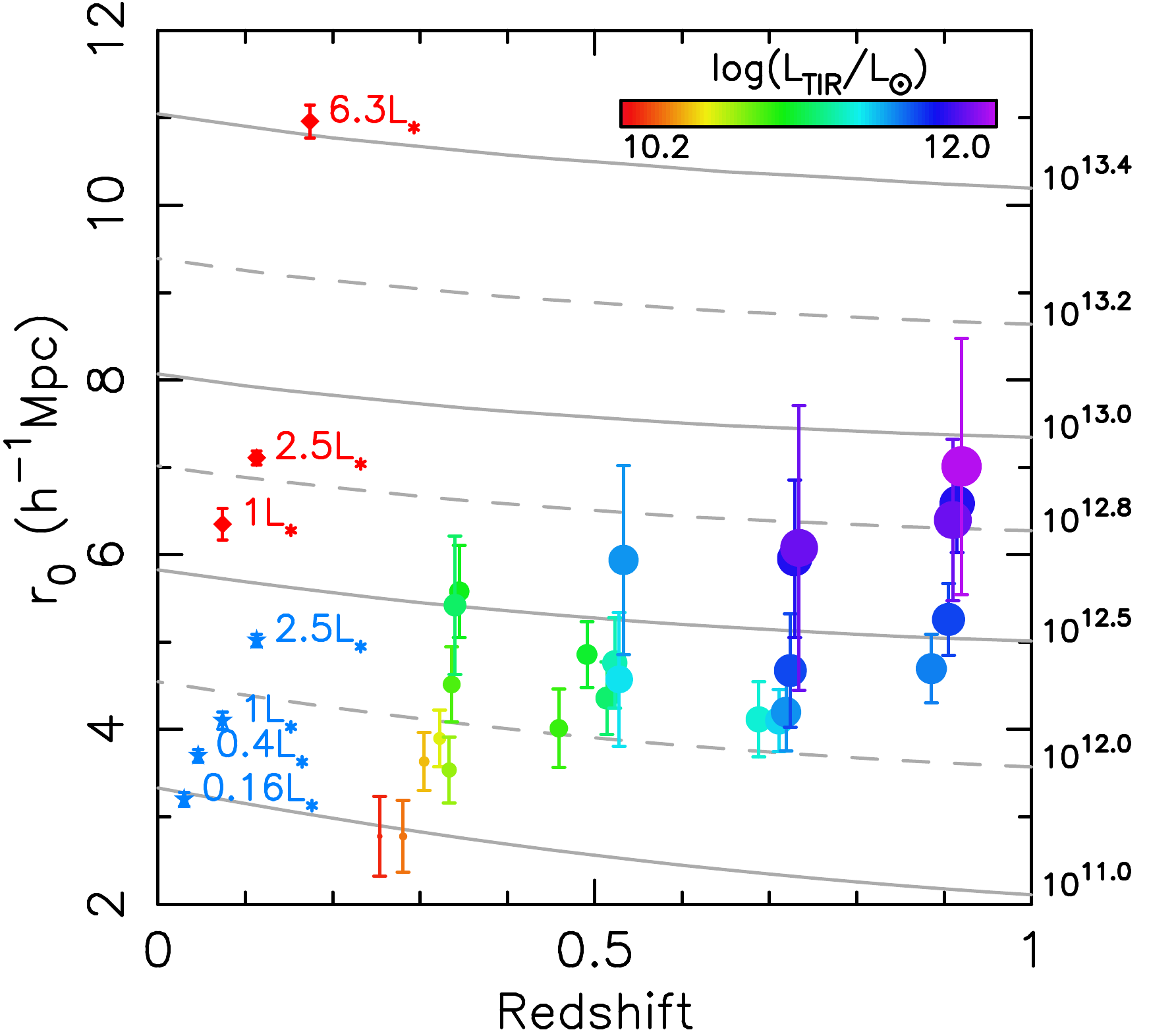}}
\caption{The clustering of star forming galaxies with a fixed $24~\mu m$ flux limit (Left) and of $L_{TIR}$ selected star forming galaxies (right).  The lines show the typical evolution of the clustering of dark matter halos, using the mean halo mass growth rates from \citet{fakho10}.  These lines can be used to connect galaxy samples to their descendants.  Halo masses are shown at $z=1$ with units of $~h^{-1}M_\odot$.  The size and color of data points for the $L_{TIR}$ selected samples are scaled linearly by $\log(L_{TIR})$.  Data points at low redshift show the clustering of red (diamonds) and blue (stars), luminosity selected galaxy samples from \citet{zehav11}.  Their $0.16L_*$ and $0.4L_*$ red galaxy samples have been removed for clarity, but both have $r_0\approx7~h^{-1}$Mpc.  This model shows that most star forming galaxies at $z<1$ are typically still forming stars in the local universe, but those with the highest SFRs at $0.6<z<1.0$ are progenitors of early-type galaxies.  Data points with the same mean redshift have been slightly offset for clarity.}
\label{fig:evoln}
\end{center}
\end{figure*}

Figure \ref{fig:evoln} shows the clustering of our star forming galaxies, and the clustering of local red and blue galaxies from \citet{zehav11} overlaid with lines showing the mean evolution of the clustering of dark matter halos.  We conclude that the majority of star forming galaxies at $z<1.0$ are star forming progenitors of $L\lesssim2.5L_*$ blue galaxies in the local universe.  However, star forming galaxies with the highest SFRs ($L_{TIR}\gtrsim10^{11.7}~L_\odot$) at $0.6<z<1.0$ are typically star forming progenitors of early-type galaxies in the local universe in the denser environments, $M_{halo}\sim10^{13}~h^{-1}M_\odot$, that typically host groups \citep{tekol13,kauff04,wang13}.  Since the bulk of star formation in early-type galaxies occurred at $z>1$ \citep[e.g.][]{franc98,renzi06}, these must be progenitors of early-type galaxies which are undergoing a final burst of star formation.

\section{Discussion}

The measured correlation length of the entire star forming galaxy sample was $r_0=(3.34\pm0.43)~h^{-1}$Mpc.  This relatively low clustering is consistent with the low redshift, blue galaxy samples of \citet{zehav11} with L$\approx0.4\rm{L_*}$, and lower than that of red galaxy samples at $z<1$ \citep{brown08,coil08,zehav11}.

For our $F_{24~\mu\rm{m}}=0.223$~mJy flux limited samples, we observe a steady increase in correlation length with redshift, from $r_0=(3.18\pm0.44)~h^{-1}$Mpc at $0.2<z<0.4$ to $r_0=(5.86\pm0.66)~h^{-1}$Mpc at $0.8<z<1.0$.  This is primarily a clustering dependence on MIR luminosity.  Due to our $24~\mu$m flux limit, we observe less luminous objects in the lower redshift samples, but only the most luminous $24~\mu$m sources at higher redshift.  When we increased the $24~\mu$m flux density limit to $0.4$~mJy we observed an increase in $r_0$ and halo mass at all redshifts.

We see a clear dependence on $L_{TIR}$  for the correlation lengths at all redshifts, where galaxies with a larger $L_{TIR}$ (hence higher SFRs) are preferentially found in higher mass halos.  A clustering dependence on optical and NIR luminosity has been known to exist for some time \citep{norbe01,baldr04,brown08,mccra08,waddi07,zehav11}.  A clustering dependence on MIR luminosity was shown to exist in MIR galaxies at $z\approx2$ by \citet{brodw08}, and we show that this dependence also applies at $z<1$.

\begin{figure}
\begin{center}
\includegraphics[width=\columnwidth]{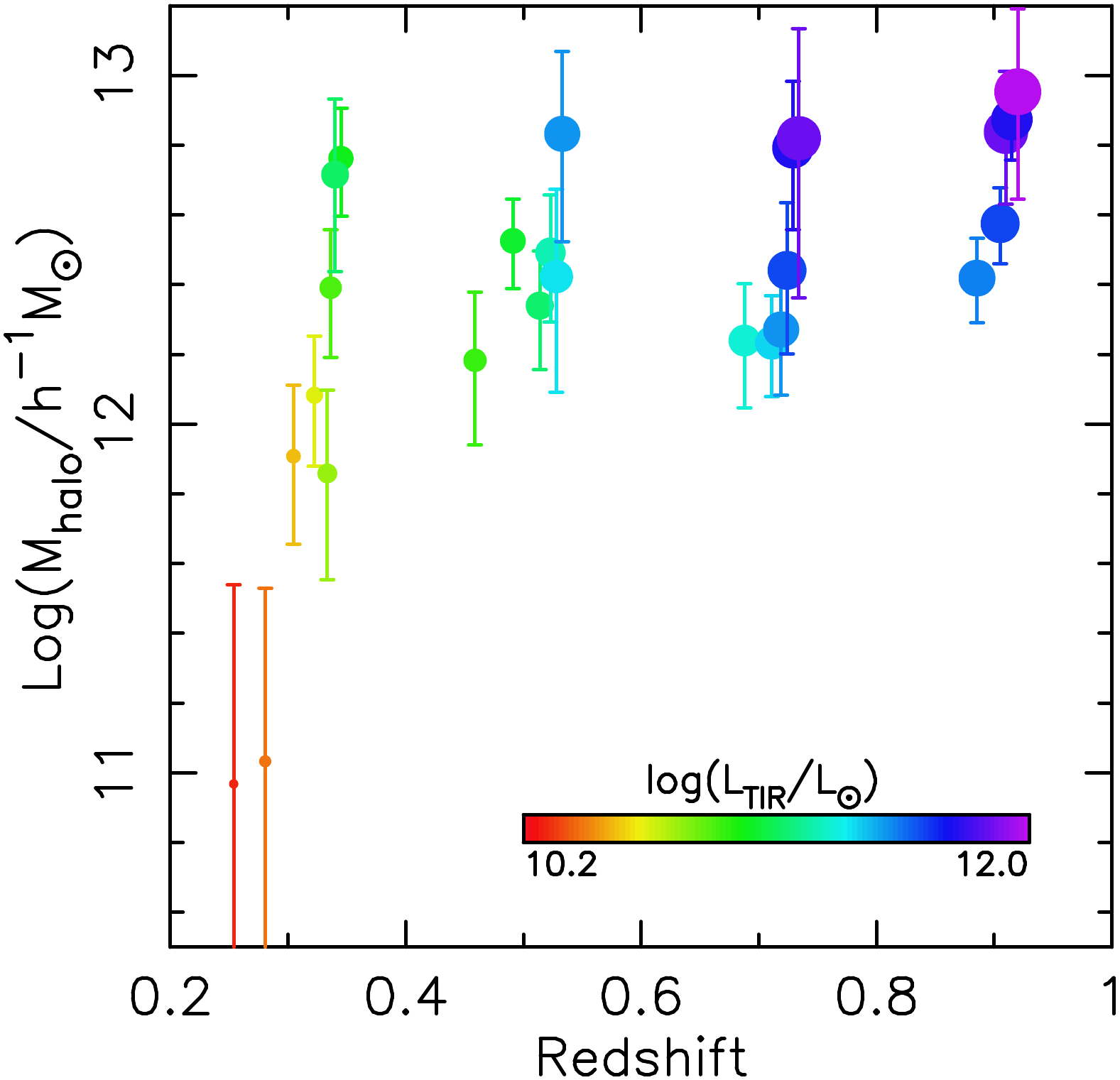}
\caption{The inferred halo masses for $L_{TIR}$ selected samples as a function of the mean redshift of each sample.  The size and color of data points are scaled linearly by $\log(L_{TIR})$.  For a fixed halo mass, the $L_{TIR}$ (and therefore the SFR) of their resident galaxies increases with redshift.  The samples with the highest SFRs at each redshift all typically reside within the same mass dark matter halos, $M_{halo}\sim10^{12.9}~h^{-1}M_\odot$.  This is consistent with a transitional halo mass where star formation is largely  truncated.  Data points with the same mean redshift have been slightly offset for clarity.}
\label{fig:Mhalovsz}
\end{center}
\end{figure}

Figure \ref{fig:Mhalovsz} shows halo mass as a function of redshift for $L_{TIR}$ selected samples, which clearly shows the $L_{TIR}$ dependence on halo mass.  However, the most luminous star forming galaxies reside within the same mass dark matter halos, $M_{halo}\sim10^{12.9}~h^{-1}M_\odot$ at all redshifts.  Even though our lowest and highest redshift samples have SFRs differing by more than an order of magnitude they reside in similar mass halos.  This complements the work of \citet{brown08}, who found that red galaxies at $z<1$ are constrained by a minimum halo mass of $\approx10^{12}~h^{-1}M_\odot$, and \citet{hartl13} who found that passive galaxies at $z<4$ are typically within halos $\geqslant10^{12.7}~h^{-1}M_\odot$.  These results are consistent with a transition region in halo mass where star formation is truncated in galaxies, however it is also possible that star forming galaxies with the highest star formation rates do reside within halos $>10^{13}~h^{-1}M_\odot$, but due to their low space density our survey volume is not large enough to measure this.

\begin{figure*}
\begin{center}
\resizebox{2\columnwidth}{!}{\plottwo{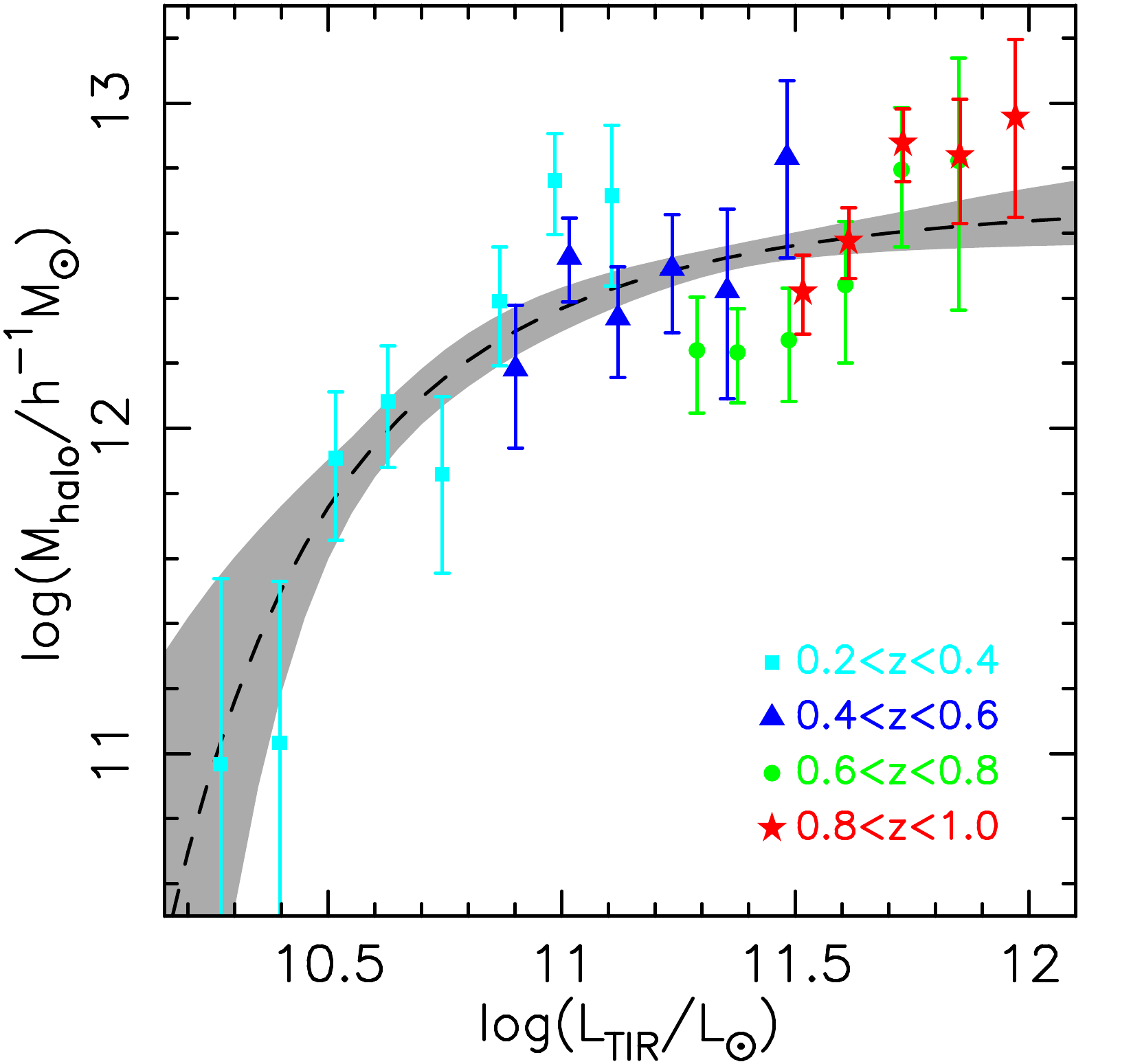}{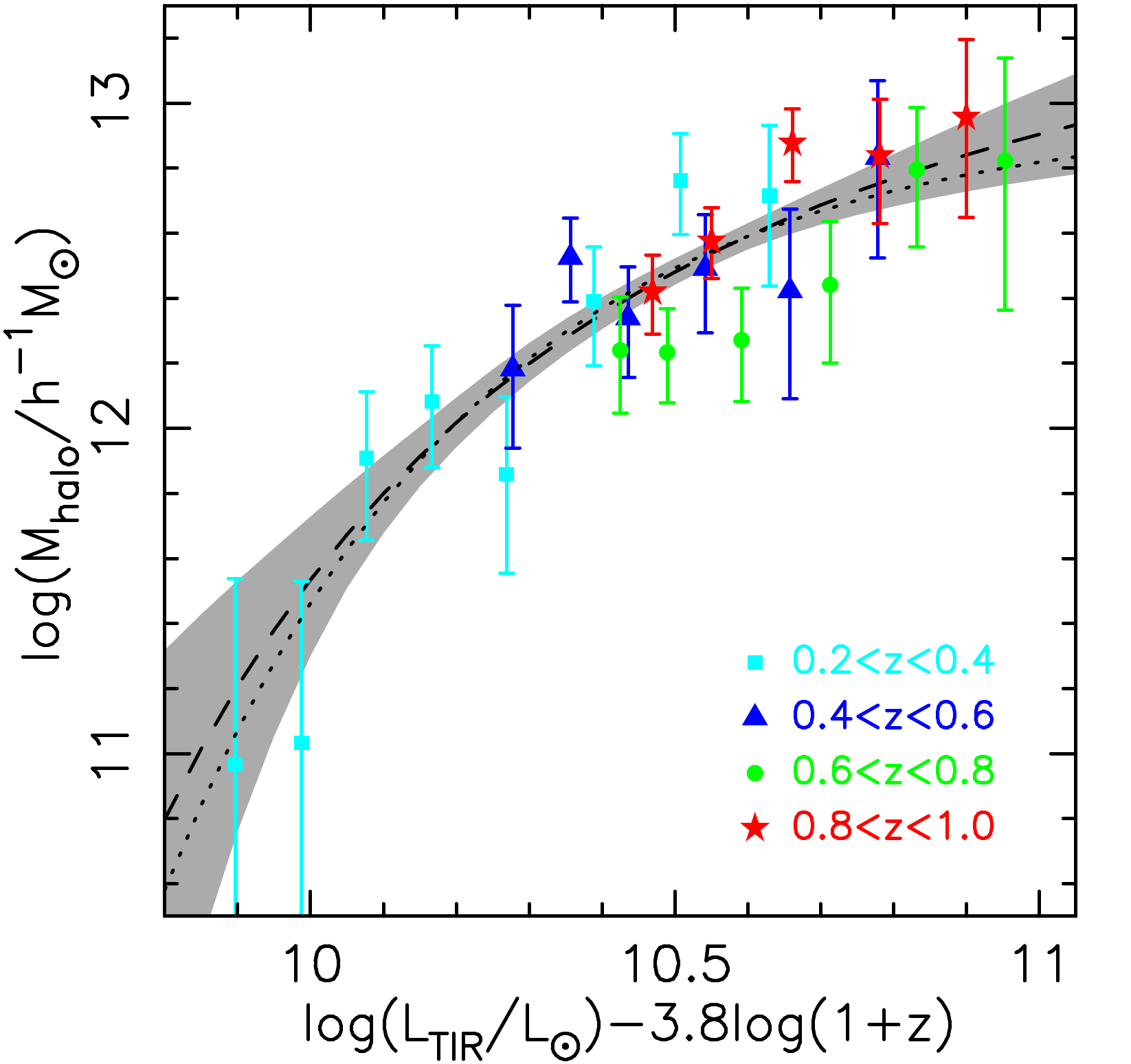}}
\caption{Typical halo mass of star forming galaxies as a function of $L_{TIR}$ (left), and as a function of $L_{TIR}$ corrected for redshift evolution (right).  There is a clear correlation between $L_{TIR}$ and typical halo mass at all redshifts, but this correlation is offset at each redshift.  When we scale $L_{TIR}$ by the evolution of the of $L_*$ in the TIR luminosity function \citep{leflo05,magne13,patel13} then the scatter in the $M_{halo}$--$L_{TIR}$ relation is reduced.  We fit a model which allowed for both a power law and an asymptotic function and find that the asymptotic function is clearly favoured, while power law fits are excluded.  The shaded regions highlights all fits within $1\sigma$ of the best.  There is a better fit to the data corrected for $L_{TIR}$ evolution, but both fits show that typical halo masses are $M_{halo}\approx10^{13}~h^{-1}M_\odot$ for galaxies with the highest SFRs, consistent with a transitional halo mass where star formation is largely truncated.  The dotted line in the right panel is the best fit exponential function, which had the lowest $\chi^2/dof$ of all models.}
\label{fig:MhalovsLIR}
\end{center}
\end{figure*}

Figure \ref{fig:MhalovsLIR} shows the typical halo mass of our star forming galaxies as a function of $L_{TIR}$.  There is a clear $M_{halo}$--$L_{TIR}$ relation, but it is offset at different redshifts.  The bright end of the IR luminosity function shifts to lower $L_{TIR}$ by $\sim1.3$ dex from $z=1$ to the present, with $L_*$ evolving as $\sim(1+z)^{3.8}$ \citep{leflo05,magne13,patel13}.  We know that this does not correspond to the evolution of individual LIRGs, but if we assume the evolution of the $M_{halo}$--$L_{TIR}$ relation is given by the evolution of the luminosity function, then the $M_{halo}$--$L_{TIR}$ relation is tightened by scaling $L_{TIR}$ by redshift, as shown in the right panel of Figure \ref{fig:MhalovsLIR}.  We find that for star forming galaxies, halo mass increases as approximately $L_{TIR}$ to the power of $1.5$, but the relation appears to flatten, perhaps with an asymptote at $M_{halo}\approx10^{13}~h^{-1}M_\odot$.  A plausible scenario is that a transition region occurs around this halo mass, where star formation is largely truncated.  However, due to the scatter of the data points we still cannot exclude that the $M_{halo}$--$L_{TIR}$ relation continues to increase to higher halo masses and there may be ULIRGs with typical halo masses $>10^{13}~h^{-1}M_\odot$.  There is evidence of ULIRGs in clusters at $z>1$ \citep{brodw13}, and well known (albeit rarer) examples of LIRGs in clusters in the nearby Universe \citep{frase14}.  \citet{weinm06} show that the fraction of central late-type galaxies in SDSS decreases very rapidly beyond halo masses of $10^{13}~h^{-1}M_\odot$, but is still greater than zero, and this fraction may increase with redshift.  We would need a larger sample to constrain the $M_{halo}$--$L_{TIR}$ relation at $L_{TIR}>10^{12}~L_\odot$ due to the low space density of ULIRGs.

If $M_{halo}\approx10^{13}~h^{-1}M_\odot$ does correspond to a transitional halo mass where star formation is largely truncated, then an asymptoting function should be preferred by the data in Figure \ref{fig:MhalovsLIR}.  To quantify this we fit the data with the functional form $\log(M_{halo})=A[\log(L_{TIR})-8.5]^B+C$, where $A$, $B$ and $C$ are the free parameters.  This functional form allows for both a power law or an asymptoting function.  The best fit parameters to the scaled $L_{TIR}$ are $A=-5.0^{+4.2}_{-1.0}$,  $B=-2.3^{+1.9}_{-1.9}$, and $C=13.5^{+5.4}_{-0.5}$, with $\chi^2=23.86$, which has much less scatter than when we fit to the unscaled $L_{TIR}$ data, with $\chi^2=34.22$.  These fits are shown as the dashed lines in Figure \ref{fig:MhalovsLIR}, where the shaded regions highlight all fits within $1\sigma$ of the best fit.  The data clearly favours an asymptoting function, with a forced a power law fit to the scaled $L_{TIR}$ having $\chi^2=27.92$.  We obtain an even tighter relation if we fit a simpler 2-parameter asymptoting model of $\log(M_{halo})=A^\prime{}-10^{B^\prime{}-\log(L_{TIR})}$.  We find $A^\prime=12.97^{+0.08}_{-0.09}$ and $B^\prime=10.18^{+0.07}_{-0.08}$, which has $\chi^2/dof=1.06$, lower than the $\chi^2/dof=1.09$ obtained for the 3-parameter model.  Both of these asymptoting models suggest that the star forming galaxies with the highest SFRs have a typical halo mass of $\sim10^{13}~h^{-1}M_\odot$, and above this mass star formation is largely truncated.

If we are seeing a transitional halo mass where star formation is truncated, then this mass helps to constrain the mechanisms responsible for quenching star formation in the densest environments.  Virial shock heating is one such mechanism that depends on halo mass.  This occurs when infalling gas is shock heated to the viral temperature of the halo, so cold gas flows can no longer efficiently feed the resident galaxies.  Simulations show that the critical mass threshold for a shock at the virial radius occurs at approximately $10^{12}~M_{\odot}-10^{12.5}~M_{\odot}$ at $z<1$ \citep{dek06,cen11}.

AGN feedback may also occur at these halo mass scales.  The ``quasar-mode'' AGN model \citep[e.g.][]{wyi03,hop06,croto06} shows that AGN can output enough energy into their host galaxy to truncate star formation.  QSOs from $0.3 < z < 2.2$ reside within a minimum halo mass of $5\times10^{12}~M_{\odot}$ \citep{croom05,farra06}, similar to the transitional halo mass we observe.  Feedback from low-accretion rate AGN can also prevent shock heated gas from cooling, preventing star formation \citep[e.g.][]{croto06,sijac06,hop06c}.  The ``radio-mode'' AGN fraction is a strong function of galaxy mass \citep[e.g.][]{sad89,best05,dek06,brown11,pimbb13}, so is more likely in higher mass halos.  \citet{coil09} find that X-ray AGN at $z\sim1$ have a clustering amplitude of $r_0=5.95\pm0.90~h^{-1}$Mpc, similar to that of passive galaxies and green valley galaxies, but also similar to our galaxies with the highest SFRs at this redshift.  Both virial shock heating and AGN feedback suggest a transitional halo mass of $\sim10^{12.5}M_\odot$, lower than the transitional mass of $\sim10^{13.0}~M_\odot$ that we find, so if any of these mechanisms is the dominant mode of truncating star formation then there cannot be a simple deterministic cut-off in star formation precisely at $10^{12.5}~M_\odot$.

It is also evident from Figure \ref{fig:Mhalovsz} that for a fixed halo mass there is an increase in $L_{TIR}$, and hence an increase in SFR with redshift, consistent with \citet{wang13}.  Since the galaxy--halo mass relation changes very little at $z<1$ \citep[e.g.][]{brown08,leaut12,hopki13,wang13}, these halos most likely contain similar mass galaxies with specific star formation rates (SSFRs) increasing with redshift, as found by \citet{alber14}.  This is consistent with the ``downsizing'' phenomenon, where the bulk of star formation is occurring in progressively lower mass galaxies with decreasing redshift \citep{cowie96}.

The results of previous MIR clustering studies with smaller samples and volumes compare well with ours.  \citet{magli08} measured the clustering of $F_{24~\mu m}>0.4$~mJy sources from the $0.7\rm{~deg}^2$ XMM-LSS field.  Their lower redshift sample contained 350 sources with photometric redshifts of $0.6<z<1.2$ and measured a correlation length of $r_0=5.95^{+1.1}_{-1.3}~h^{-1}$Mpc.  The uncertainties in their results are estimated as Poisson random errors in the galaxy pair counts, which only provide a lower limit to the actual uncertainties.  They estimate that $\sim40\%$ of their sources are AGN, so this is not a pure star forming sample.  Their correlation length is consistent with our overlapping samples, with $r_0=(5.26\pm0.77)~h^{-1}$Mpc at $0.6<z<1.0$ and $r_0=(7.20\pm1.18)~h^{-1}$Mpc at $0.8<z<1.0$, however our sample and field size are more than an order of magnitude larger, making our results far less prone to statistical uncertainties and cosmic variance.

\citet{gilli07} measured the clustering of $\sim1300$ $F_{24~\mu\rm{m}}>20~\mu$Jy sources in the GOODS fields with a mean redshift of $z\sim0.7$.  They measure a correlation length of $r_0=(4.03\pm0.38)~h^{-1}$Mpc, and find that this increases to $r_0=(5.14\pm0.76)~h^{-1}$Mpc for ULIRGs.  While these results agree with our similar samples within experimental uncertainties, their sample spans the broad redshift range [0.1,1.4], which encompasses half the age of the universe, possibly including many different evolutionary stages of star forming galaxies.

\citet{stari12} measured the clustering of $24~\mu$m galaxies in the SWIRE Lockman Hole field, with 14822 $F_{24~\mu\rm{m}}>310~\mu$Jy sources.  They measure a correlation length of $(4.98\pm0.28)~h^{-1}$Mpc, also in agreement with our results, but they have no redshifts for their sample, so they use a color cut to restrict the sample to broad redshift range similar to that of \citet{gilli07}.  The model fits to the angular correlation function of \citet{stari12} predict an upturn at scales smaller than they are able to measure due to source confusion.  We measure the correlation function for 3 bins below their limit, down to angular scales of $10^{\prime\prime}$, and do indeed see an upturn in the correlation function.  We inspected close pairs in both the $24~\mu$m and $I$-band imaging to confirm these were distinct galaxy pairs, and not split sources giving a false excess of small scale clustering.  This excess in the correlation function at small scales is most likely due to pairs of star forming galaxies within the same dark matter halos.  The inclusion of these data points in our power law fits explains why we obtain slightly steeper fits than previous studies.

\begin{figure*}
\begin{center}
\includegraphics[width=1.8\columnwidth]{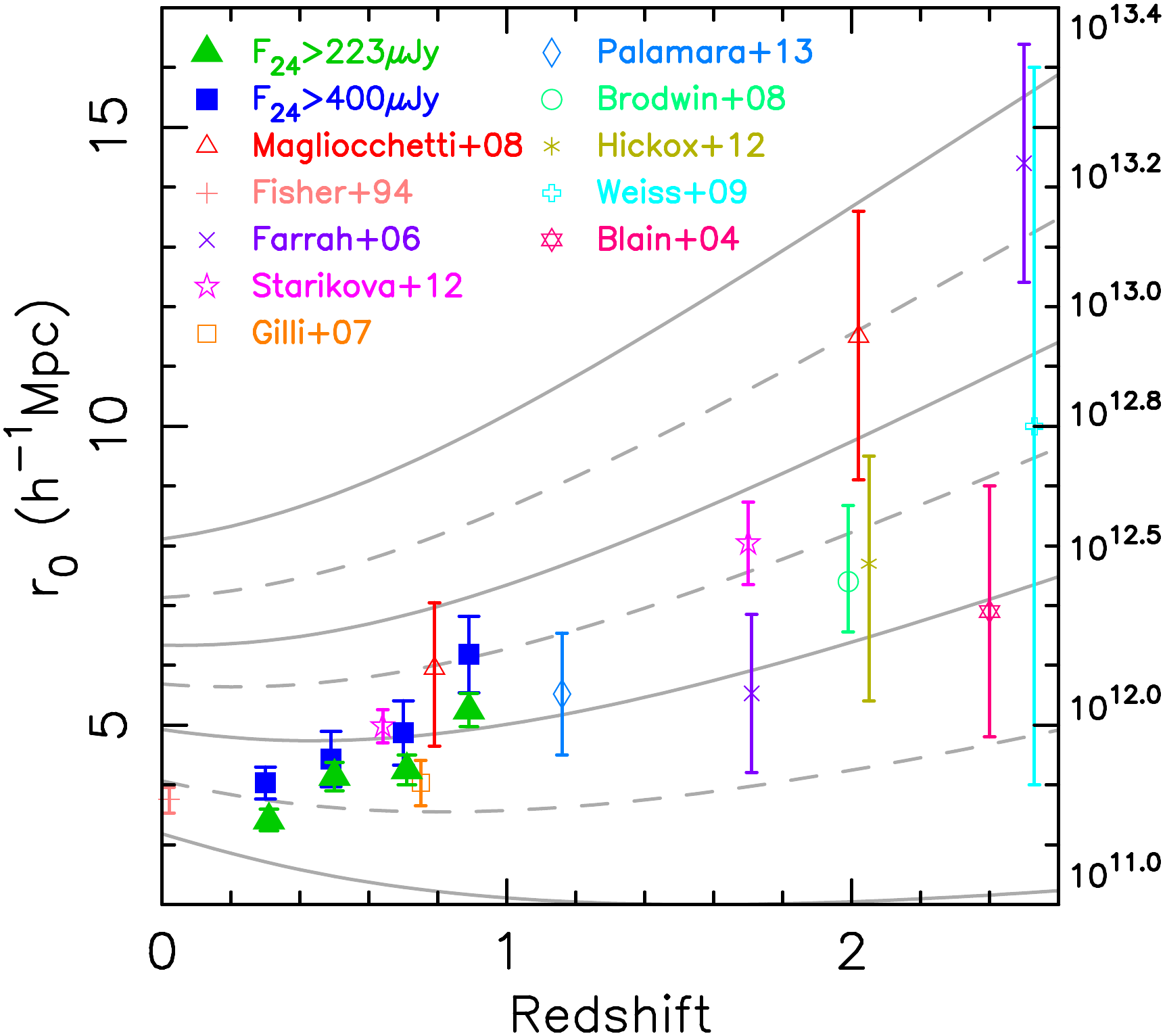}
\caption{A comparison of the clustering of MIR and FIR galaxy samples.  The lines show the clustering for fixed mass dark matter halos, with halo masses shown to the right in $h^{-1}M_\odot$.  These lines do not represent the evolution of galaxy clustering, since dark matter halos gain mass through mergers.  We see an increase in correlation length and halo mass with redshift, but this is primarily an IR luminosity dependence, since at a fixed flux limit we see galaxies with progressively greater IR luminosities (and SFRs) at higher redshift.  The low redshift $r_0$ value of \citet{farra06} has been revised down by a factor of 1.7, because their redshift distribution was later found to be narrower than expected by almost a factor of 3.  Some data points have been offset slightly in redshift for clarity.}
\label{fig:r0allMIR}
\end{center}
\end{figure*}

Figure \ref{fig:r0allMIR} shows a comparison with previous MIR \citep{fishe94a,farra06,magli08,gilli07,brodw08,stari12,palam13} and FIR/sub-mm \citep{blain04,weiss09,hicko12} galaxy clustering results, overlaid with lines of fixed halo mass so halo masses can be compared.  The low redshift $r_0$ value of \citet{farra06} has been revised down by a factor of 1.7, since in \citet{farra08} they show that the redshift distribution of their sample was narrower by almost a factor of 3 than they initially estimated from photometric redshifts.  All correlation lengths at $z<2$ are consistent with a transitional halo mass of $M_{halo}\approx10^{13.0}~h^{-1}M_\odot$ where star formation is truncated in galaxies.  The samples at $z>2$ suggest star formation was occurring in higher mass halos, however the uncertainties in these estimates are large, and likely to be underestimates in many cases, due to the assumption of Poisson errors.  These samples also do not all have AGN removed, and contain galaxies over broad redshift ranges, so it is possible that the halo mass threshold we find for star forming galaxies at $z<1$ does extend to higher redshifts.  Larger MIR surveys at $z>1$ with well constrained redshifts would be required to confirm such a trend.

Future HOD analyses would give a more precise description of how star forming galaxies are distributed within dark matter halos as a function of halo mass and redshift.  This would shed light on the dominant processes responsible for regulating star formation within galaxies.  If galaxy SFR at a fixed halo mass has a strong dependence on the number and distribution of satellite galaxies, then mergers are playing a significant role in boosting SFRs.  More robust clustering measurements for star forming galaxies at high redshift would allow star forming galaxies at earlier epochs to be connected to local populations. This would confirm whether giant elliptical galaxies could indeed be formed by truncation of star formation in massive galaxies, or alternatively it would show that such massive galaxies can only be formed by hierarchical growth.

\section{Summary}
We measured the clustering and dark matter halo masses of $24~\mu$m selected star forming galaxies at $0.2<z<1.0$.  Our sample comprises 22553 star forming galaxies from $8.42\rm{~deg}^2$ of the Bo\"otes field.  This is a larger sample size and field area than all previous MIR clustering studies at similar redshifts.  Selection based on $24~\mu$m emission allows us to observe galaxies with the highest SFRs, which can be difficult when selecting star forming samples with optical data, because of varying levels of dust obscuration.  Our main results are:

\begin{itemize}

\item We find that the galaxies with the highest SFRs have optical colors which are redder than typical blue cloud galaxies, and many reside within the green valley, consistent with \citet{weine05} and \citet{bell05}.  Examination of the axis ratio distribution shows that red star forming galaxies are $\sim1.5$ times as likely to have an axis ratio less than 0.5 than blue star forming galaxies, so the red optical colors are due to reddening by dust within these galaxies and not by AGN contributing to the MIR emission of these galaxies.

\item We find that the measured correlation lengths and halo masses of star forming galaxies have a dependence on IR luminosity at all redshifts.  Galaxies with higher star formation rates are found in increasingly massive halos.

\item We observe relatively weak clustering of $r_0\approx3-6~h^{-1}$Mpc for most of our star forming samples at $z<1.0$.  We model the evolution of halo clustering, and conclude that the majority of star forming galaxies at $z<1.0$ are typically star forming progenitors of $L\lesssim2.5L_*$ blue galaxies in the local universe, while star forming galaxies with the highest SFRs ($L_{TIR}\gtrsim10^{11.7}~L_\odot$) at $0.6<z<1.0$ are typically star forming progenitors of early-type galaxies, in denser group environments.

\item While many of our star forming galaxies at $z\approx1$ are typically progenitors of early-type galaxies, they are not the progenitors of giant elliptical galaxies ($L>2.5L_*$), so these must be formed either by mergers or by the truncation of star formation in even more massive galaxies at $z>1$.

\item The samples with the highest $L_{TIR}$ (hence the highest SFRs) at each redshift typically reside within halos with $M_{halo}\approx10^{12.9}~h^{-1}M_\odot$.  This is consistent with a transition region in halo mass where star formation is largely truncated, however our data do not exclude the possibility of star forming galaxies within more massive halos.

\item For a constant halo mass, the SFRs of the resident galaxies increases with redshift.  This is not unexpected, as \cite{alber14} show that SFRs of galaxies with a fixed stellar mass increase with redshift, and there is little evolution of the galaxy-halo mass relation at $z<1$ \citep[e.g.][]{leaut12,hopki13}.  This is consistent with the observed ``downsizing'' phenomenon, where the bulk of star formation is occurring in progressively lower mass galaxies \citep{cowie96} with decreasing redshift.

\end{itemize}

\begin{acknowledgments}

We thank the anonymous referee for their helpful suggestions and constructive criticism.  This work is based in part on observations made with the Spitzer Space Telescope, Spitzer/IRAC and Spitzer/MIPS, which is operated by the Jet Propulsion Laboratory, California Institute of Technology, under a contract with NASA. This work is based in part on observations made with the Kitt Peak National Observatory (KPNO). This research was supported by the National Optical Astronomy Observatory, which is operated by the Association of Universities for Research in Astronomy (AURA), Inc., under a cooperative agreement with the National Science Foundation. We thank our colleagues on the NDWFS, SDWFS and MAGES teams.  T.D. acknowledges support from an Australian Postgraduate Award (APA) and a J.L.William postgraduate award.  M.J.I.B. acknowledges support from a Future Fellowship.  A.D. acknowledges support from the Radcliffe Institute for Advanced Study.

Funding for the SDSS and SDSS-II has been provided by the Alfred P. Sloan Foundation, the Participating Institutions, the National Science Foundation, the U.S. Department of Energy, the National Aeronautics and Space Administration, the Japanese Monbukagakusho, the Max Planck Society, and the Higher Education Funding Council for England. The SDSS Web Site is http://www.sdss.org/.

The scientific results reported in this article are based in part on observations made by the  \emph{Chandra X-ray Observatory} and published previously in cited articles.

\end{acknowledgments}

\bibliographystyle{apj}
\bibliography{24mic}

\end{document}